\documentclass[times,twocolumn,final]{elsarticle}

\pdfoutput=1

\usepackage{epstopdf}

\usepackage{cag}
\usepackage{framed,multirow}

\usepackage{amssymb}
\usepackage{latexsym}

\usepackage{url}
\usepackage{xcolor}
\definecolor{newcolor}{rgb}{.8,.349,.1}

\usepackage{subfig}
\usepackage{enumitem}

\usepackage{hyperref}

\usepackage[switch,pagewise]{lineno} 

\usepackage{algorithm}
\usepackage{algpseudocode}

\journal{Computers \& Graphics}

\begin{document}

\verso{Preprint Submitted for review}

\begin{frontmatter}

\title{OMiCroN - Oblique Multipass Hierarchy Creation while Navigating}%

\author[1]{Vin\'icius \snm{da Silva}\corref{cor1}}
\cortext[cor1]{Corresponding author: 
  Tel.: +55-21-97171-0941;}
\emailauthor{dsilva.vinicius@gmail.com}{Vin\'icius da Silva}
    
\author[2]{Claudio \snm{Esperan\c ca}}
\author[2,3]{Ricardo \snm{Marroquim}}

\address[1]{Institute for Pure and Applied Mathematics (IMPA). VISGRAF Lab. Estrada Dona Castorina, 110, Jardim Bot\^{a}nico, Rio de Janeiro - RJ, Brazil, CEP: 22460-320}
\address[2]{Federal University of Rio de Janeiro (UFRJ). Computer Graphics Lab (LCG). Cidade Universit\' aria, Centro de Tecnologia, Block H, Rio de Janeiro - RJ, Brazil, CEP: 21941-972}
\address[3]{Delft University of Technology (TU Delft). Computer Graphics and Visualization Group. Van Mourik Broekmanweg 6, Delft, The Netherlands}

\received{\today}

\begin{abstract}
Rendering large point clouds ordinarily requires building a hierarchical data structure for accessing the points that best represent the object for a given viewing frustum and level-of-detail. The building of such data structures frequently represents a large portion of the cost of the rendering pipeline both in terms of time and space complexity, especially when rendering is done for inspection purposes only. This problem has been addressed in the past by incremental construction approaches, but these either result in low quality hierarchies or in longer construction times. In this work we present OMiCroN -- Oblique Multipass Hierarchy Creation while Navigating -- which is the first algorithm capable of immediately displaying partial renders of the geometry, provided the cloud is made available sorted in Morton order. OMiCroN is fast, being capable of building the entire data structure in memory spending an amount of time that is comparable to that of just reading the cloud from disk. Thus, there is no need for storing an expensive hierarchy, nor for delaying the rendering until the whole hierarchy is read from disk. In fact, a pipeline coupling OMiCroN with an incremental sorting algorithm running in parallel can start rendering as soon as the first sorted prefix is produced, making this setup very convenient for streamed viewing. 

\end{abstract}


\end{frontmatter}


\section{Introduction}
\label{section:intro}
In recent years, improvements in acquisition devices and techniques have led to the creation of huge point cloud datasets. 
Direct rendering of such datasets must resort to indexing data structures. These are used for culling portions of the model outside the viewing frustum and for selecting representative point subsets for the portions inside it. In many use cases, the cost of building such structures is not critical, since the resulting hierarchy is stored in secondary memory so it can be reused every time a render session starts. Thus, research focusing on the quality of the render need not justify arbitrarily long preprocessing times (e.g. \cite{Rusinkiewicz2000, Wimmer2006}). In other cases, shortening the time to produce the hierarchy is deemed worthwhile, at the expense of achieving slightly worse balance or render quality. This is particularly useful for applications that must render the point cloud and perform additional tasks or that must handle dynamic data (e.g. collision detection~\cite{Klein2004}). No published research, to the best of our knowledge, has yet reported a means for rendering point clouds \emph{before} the hierarchy is completely available. 

\paragraph{Bottom-up hierarchy building} Strategies for building point cloud hierarchies can be divided into three classes: incremental, top-down and bottom-up. Incremental strategies consist of sequentially inserting points into an incomplete hierarchy. The main limitation of this strategy is that the quality of results are ultimately dependent on the insertion order~\cite{Ericson2004,Bittner2015}. Better quality hierarchies require examining the whole data beforehand. Top-down strategies work by partitioning the input in increasingly smaller groups. On the other hand,  bottom-up strategies join small collections of close points into increasingly larger groups. One simple way of producing a sequence of points that in general lie close to each other is to sort them according to some 3D space-filling curve such as that defined by the \textit{Morton order}, used to organize nodes in octrees. 

\paragraph{OMiCroN} In this paper we introduce OMiCroN (Oblique Multipass Hierarchy Creation while Navigating), a new take on the problem of shortening the delay between point cloud acquisition and its visualization. Its central idea -- and main contribution -- is to build the hierarchy in memory while allowing a synchronous inspection of all data already loaded. It is important to stress that performing both tasks in parallel involves solving non-trivial synchronization issues. OMiCroN circumvents most of these by combining Morton code ordering, bottom-up construction and the concept of \textit{oblique cuts}, where the renderable parts of the model are clearly separated from the non-renderable parts by a single delimiting Morton code.

\begin{figure}[htb!]
  \centering
    \includegraphics[width=1.0\columnwidth]{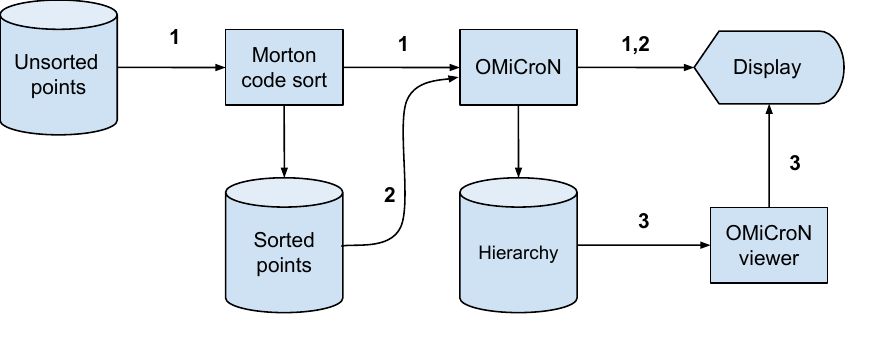}
  \caption{The standard OMiCroN pipeline (arrows labeled \textbf{1}) permits inspecting a raw point cloud where the first images are produced just after an incremental Morton code sorter outputs the first points. If the sorted points are already available, OMiCrON starts rendering immediately (arrows labeled \textbf{2}). The hierarchy computed by OMiCroN can be flushed to disk for later reuse. This is the traditional pipeline (arrows labeled \textbf{3}), where rendering starts after the hierarchy is built.}
  \label{fig:pipelines}
\end{figure}

\paragraph{Use cases} The fact that OMiCroN only requires that the input point cloud be ordered by Morton Code allows it to be deployed in several ways. For instance, a pipeline can be built where an unordered point cloud is fed into an incremental sorter and, as soon as the sorted points are produced, they are fed into OMiCroN. Alternatively, one can use a batch sorter to produce a file containing the ordered points file, which is later read by OMiCroN. In both cases, the hierarchy produced by OMiCroN can be stored as a file for later reuse (see Figure~\ref{fig:pipelines}).

\paragraph{Contributions} The technical contributions of this work are:

\begin{itemize}
  \item introduces the concept of Hierarchy Oblique Cuts, that allows parallel data sorting, spatial hierarchy construction and rendering; 
  \item restricts the preprocessing of input data to a very fast and flexible Morton code based partial sort;
  \item allows for on-the-fly Octree construction for large point clouds;
  \item following the Morton Order, renders full detail data from the very beginning as a consequence of bottom-up hierarchy construction; 
  \item provides immediate visual feedback of the hierarchy creation process.
\end{itemize}

This paper is organized in the following manner. In Section~\ref{section:background} we present the necessary background for describing OMiCroN.
In Section~\ref{section:related} the related work is presented.
In Section~\ref{section:overview} we give an overview of our method, while the two central concepts of Hierarchy Oblique Cuts and Oblique Hierarchy Cut Fronts are described in details in Sections~\ref{section:cuts} and~\ref{section:front}, respectively. 
In Section~\ref{section:omicron} we present the parallel version of the OMiCroN algorithm, describing a proof-of-concept application for processing and rendering large point clouds.
In Section~\ref{section:comparisons} we describe the experiments to measure the preprocessing, rendering and memory efficiency of the algorithm.
Finally, conclusions, limitations and future work directions are presented in Section~\ref{sec:discussions}.

\section{Background}
\label{section:background}

Our work depends on three major concepts: Morton Order; Hierarchical Spatial Data Structures; and Rendering Fronts. The theory behind them is summarized in this section.

\begin{figure}[htb!]
  \centering
  \subfloat[Z-order curve.]{
    \includegraphics[width=.4\columnwidth]{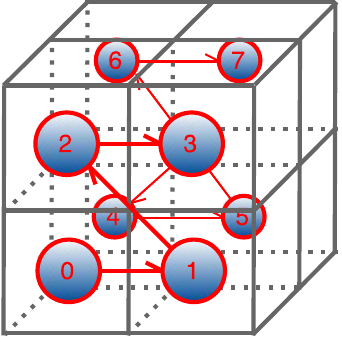}
    \label{fig:morton_order}
  }
  \hfil
  \subfloat[Relationship between Morton order and grid coordinates.]{
    \includegraphics[width=.5\columnwidth]{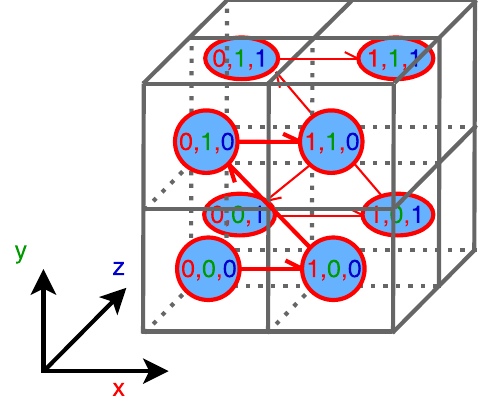}
    \label{fig:morton_coords}
  }
  \hfil
  \subfloat[Morton order and associated hierarchical representation. Order is indicated inside nodes, coordinates and Morton codes outside them.
  The Morton code for the $n$-th child of a parent node with code $x$ is $x$ concatenated with the binary (bit-interleaving) representation of $n$.  Coordinate values and interleaved bits share color. Parent code is between curly brackets and node index between square brackets. A prefix bit is used to avoid ambiguity.] {
    \includegraphics[width=1.0\columnwidth]{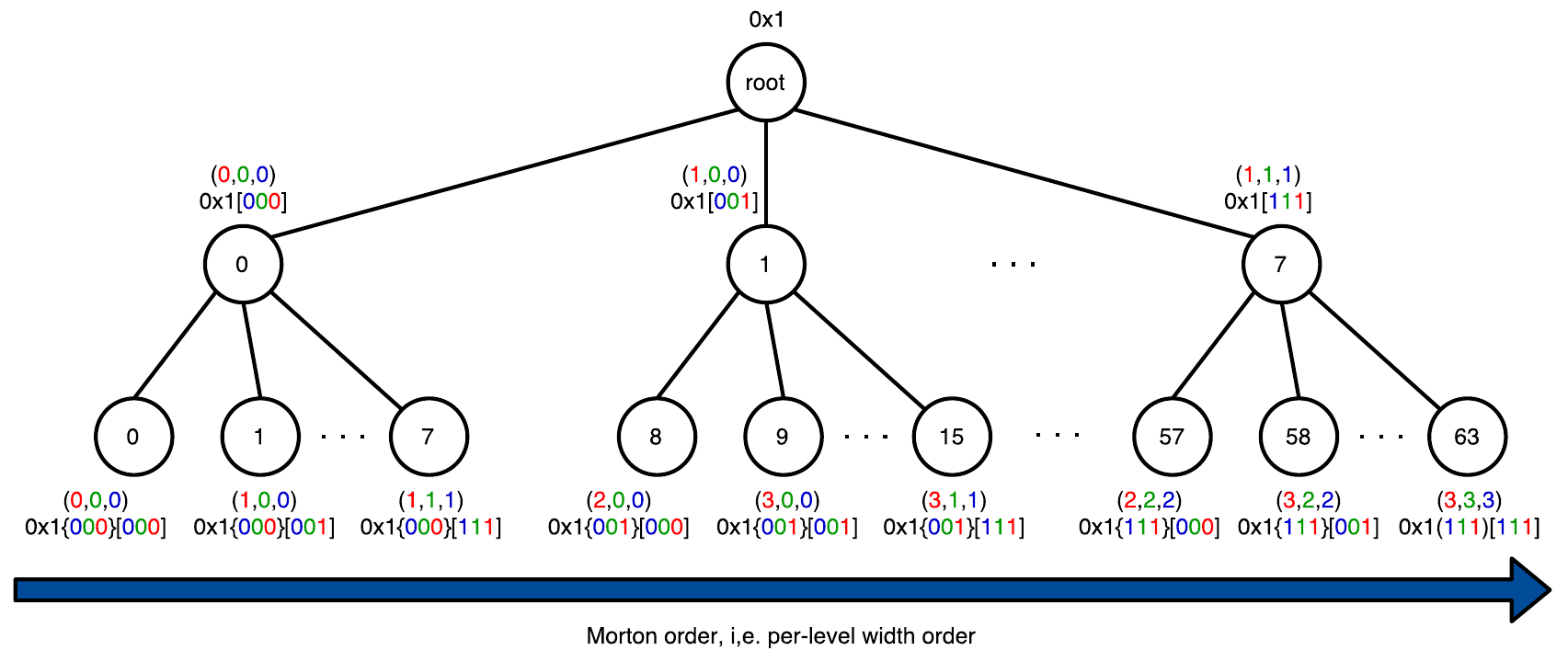}
    \label{fig:morton_tree}
  }
  \caption{Z-Order and Morton code illustrative example.}
  \label{fig:morton}
\end{figure}

\paragraph*{Morton Order} Morton~\cite{MORTON1966} proposed a linearization of 2D grids, later generalized to n-dimensional grids. It results in a z-shaped space-filling curve, called the Z-order curve.
The order in which the grid cells are visited by following this curve is called Morton order or Z-order.
The associated Morton code for each cell can be computed directly from the grid coordinates by interleaving their bits. Figure~\ref{fig:morton} illustrates the concepts above.



\paragraph*{Spatial Data Structures} Morton codes extend naturally to regular spatial subdivision schemes, thus they are usually used in conjunction with hierarchical spatial data structures such as Octrees and regular Kd-trees (Bintrees). They provide fast data culling and a direct level-of-detail structure, by mapping the n-dimensional structure to a one-dimensional list. Figure~\ref{fig:morton} illustrates an Octree with an embedded Morton code curve, and its associated hierarchical representation.

\begin{figure}[h!]
  \centering
  \subfloat[A front and operations to be performed.]{
    \includegraphics[width=.4\columnwidth]{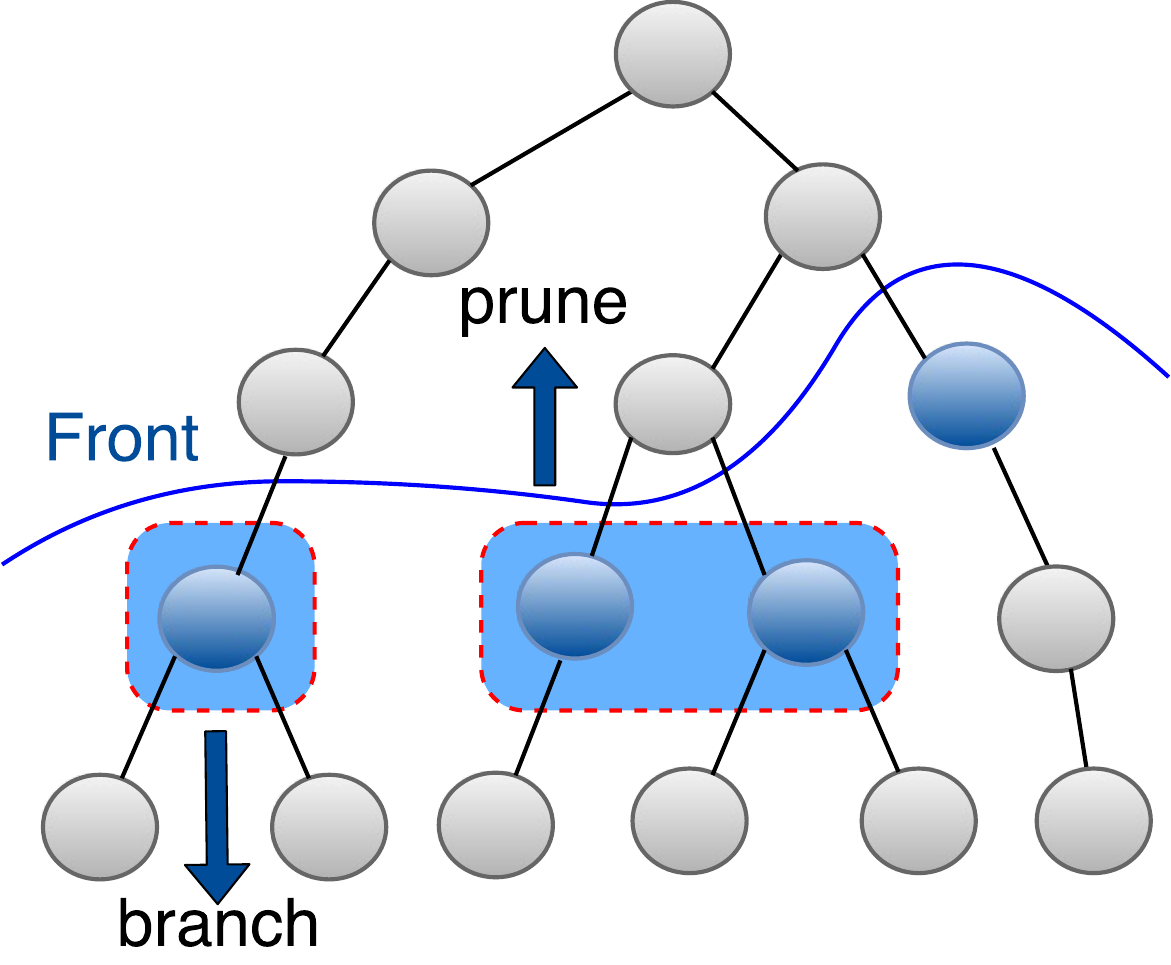}
    \label{fig:front_operators_a}
  }
  \hfil
  \subfloat[The front after the $prune$ and $branch$ operations.]{
    \includegraphics[width=.4\columnwidth]{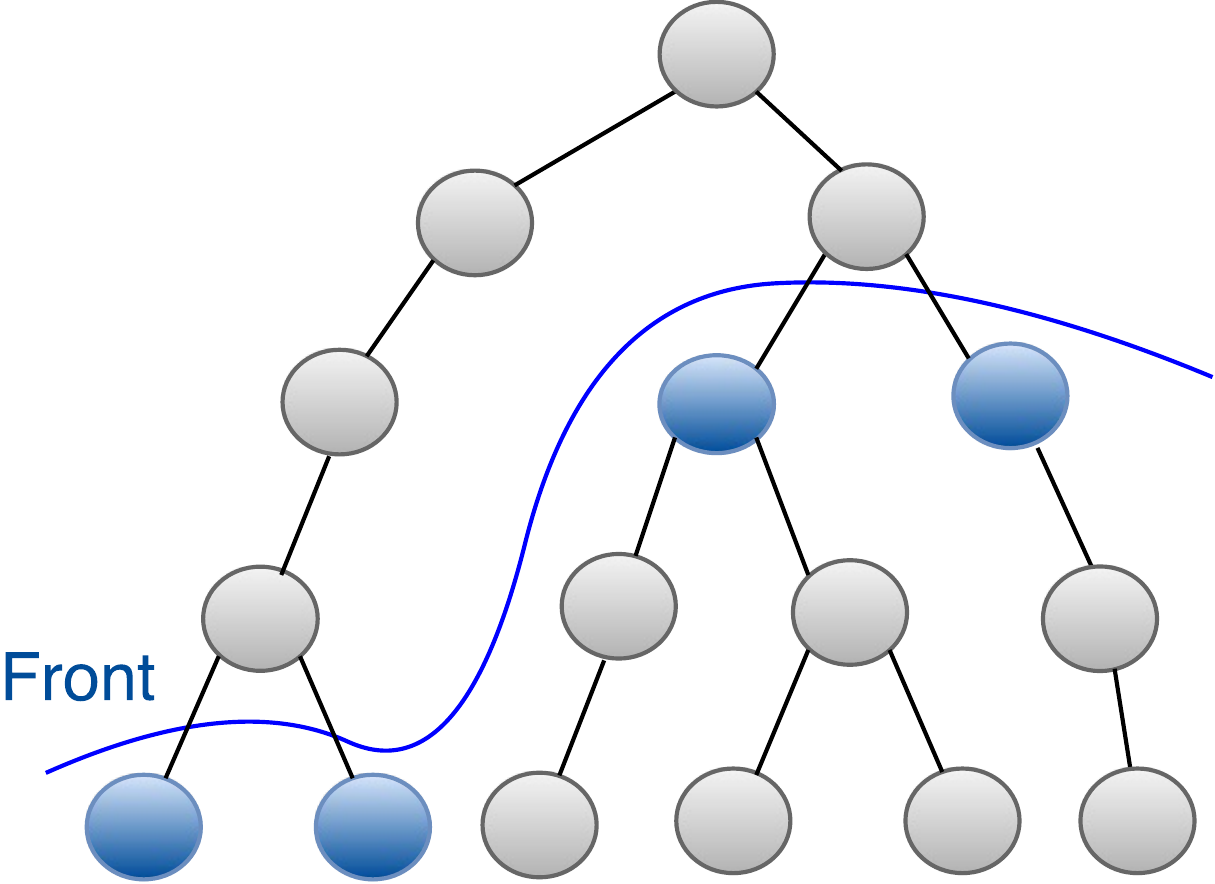}
    \label{fig:front_operators_b}
  }
  \hfil
  \caption{Rendering Front example.}
  \label{fig:front_operators}
\end{figure}

\paragraph*{Rendering Front} A Rendering Front, hence called only Front, is a structure to optimize sequential traversals of hierarchies, and has been used in many works~\cite{Klosowski1998,ehmann2001,LAUTERBACH2010,Argudo2016}. This technique explores spatial and temporal locality. Instead of starting the traversal at the root node for every new frame, it starts at the nodes where it stopped in the preceding frame. Fronts have two basic operators: \textit{prune} and \textit{branch}. The \textit{prune} operator traverses the hierarchy up, removing a group of sibling nodes from the front and inserting their parent. The \textit{branch} operator works in the opposite direction, by removing a node from
the front and inserting its children. Figure~\ref{fig:front_operators} depicts a front and the two operators.

\section{Related work}
\label{section:related}


While the use of points as rendering primitives was introduced very early in Computer Graphics \cite{Levoy1985,Grossman1998}, their widespread adoption only occurred much later, as discussed on extensive survey literature \cite{Sainz2004,Kobbelt2004,Alexa2004,Gross2006,Gross2011,Ramos2016}. Many algorithms were presented from that period on, proposing improved image quality by changes in the kernel logic, better spatial management by the use of multiresolution and LOD structures, and integration of the out-of-core paradigm, resulting in systems that can handle extremelly large point clouds. Here we focus the discussion on multiresolution and LOD structures, establishing an argument for why a stream-and-feedback-based algorithm such as OMiCroN is a desirable tool for the academy and industry.

QSplat~\cite{Rusinkiewicz2000} is the seminal reference on large point cloud rendering. It is based on an out-of-core hierarchy of bounding spheres, which is traversed to render the points. Since its main limitation is the extensive CPU usage, QSplat was followed by works focused on loading more work onto the GPU. For example, Sequential Point Trees~\cite{Dachsbacher2003} introduced adaptive rendering completely on the graphics card by defining an octree linearization that can be traversed efficiently using the GPU architecture. Other methods used approaches relying on the out-of-core paradigm, such as XSplat~\cite{Pajarola2005} and Instant Points~\cite{Wimmer2006}. XSplat proposed a paginated multiresolution point-octree hierarchy with virtual memory mapping, while Instant Points extended Sequential Point Trees by nesting linearized octrees to define an out-of-core system. Layered Point Clouds~\cite{Gobbetti2004} proposed a binary tree of precomputed object-space point cloud blocks that is traversed to adapt sample densities according to the projected size in the image. Wand et al.~\cite{Wand2007} presented an out-of-core octree-based renderer capable of editing large point clouds and Bettio et al.~\cite{Bettio2009} implemented a kd-tree-based system for network distribution, exploration and linkage of multimedia layers in large point clouds. Other works focused on parallelism using multiple machines to speed-up large model processing or to render on wall displays using triangles, points, or both \cite{Hubo2005,Correa2002,Correa2003,Goswami2010,Goswami2013}.

More recently, relatively few works have focused on further improving the rendering of large point clouds, such as the method by Lukac et al.~\cite{Lukac2014}. Instead, more effort has been concentrated on using established techniques in domains that require the visualization of large datasets as a tool for other purposes. For example, city visualization using aerial LIDAR~\cite{Gao2014,Richter2015}, sonar data visualization~\cite{Febretti2014} and, more prominently, virtual reality \cite{Potenziani2015,Tredinnick2015,Okamoto2016,Ponto2017}.

An important discussion concerns which approach best exploits parallelism when creating a hierarchy. A good way to address this question is to study GPU algorithms, which must rely on smart problem modeling to achieve maximum degree of data independency, increasing throughput in a GPU manycore environment. Karras~\cite{Karras2012} made an in-depth discussion about this subject. His major criticism of other methods is that top-down approaches achieve a low degree of parallelism at the top levels of the tree, generating underutilization of processing resources at early stages of hierarchy construction. Bottom-up methods do not suffer from this problem because the number of nodes grows exponentially with the hierarchy depth, providing sufficient data independency and a good degree of parallelism.


While the aforementioned papers present very useful and clever methods to implement or use large point cloud rendering, none of them considers presenting data to the user before the full hierarchy is created. For example, implementors of systems that use large point cloud rendering as a tool could use the visual feedback given by the algorithm in order to check if the data is presented properly, without having to wait for the full hierarchy to be available. Additionally, in environments where data transfer is a bottleneck, the input data could be transfered and the hierarchy constructed on-the-fly, instead of transferring the full hierarchy which may be several times larger.

\section{Overview}
\label{section:overview}


\begin{figure*}[!h]
  \centering
  \subfloat[Initial (possibly empty) renderable hierarchy and \textit{concatenate} operator.
  ]{
    \includegraphics[width=.2\textwidth]{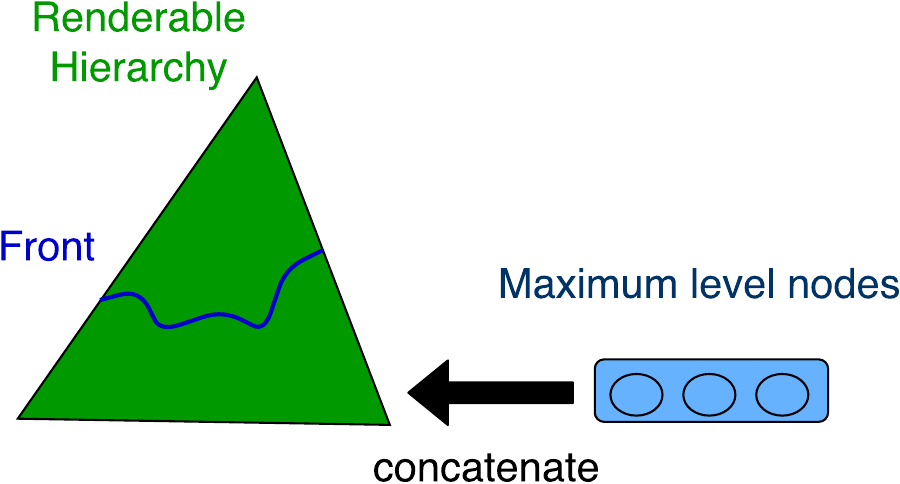}
    \label{fig:omicron_overview_a}
  }
  \hfil
  \subfloat[The \textit{fix} operator: node ancestors 
  are inserted into the hierarchy.]{
    \includegraphics[width=.2\textwidth]{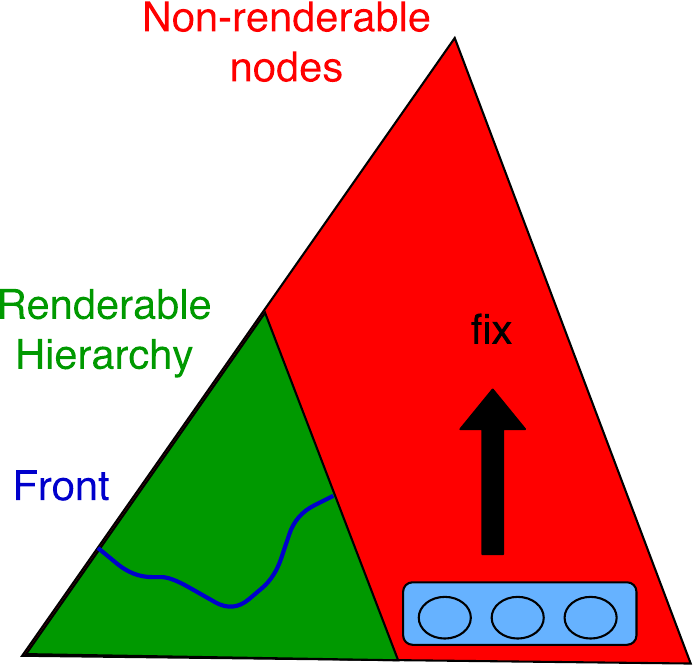}
    \label{fig:omicron_overview_b}
  }
  \hfil
  \subfloat[After the \textit{fix} operation the renderable hierarchy is expanded.]{
    \includegraphics[width=.2\textwidth]{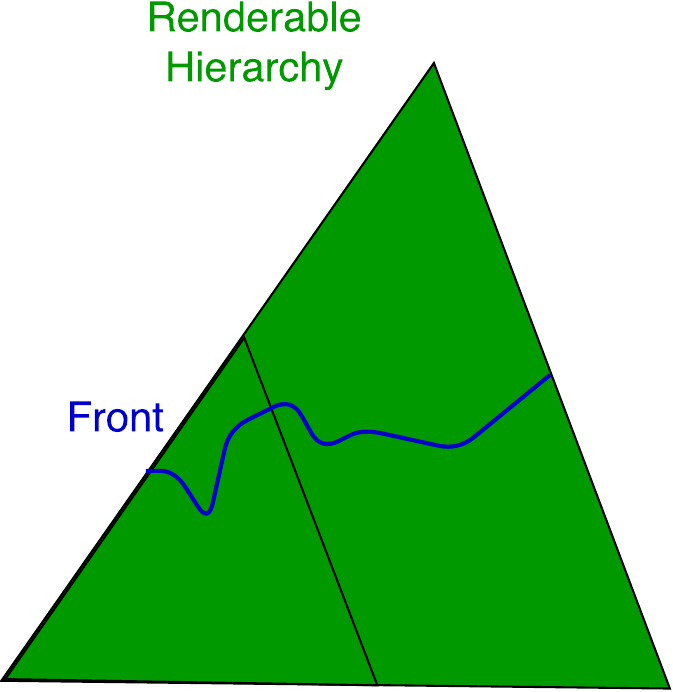}
    \label{fig:omicron_overview_c}
  }
  \hfil
  \caption{OMiCroN overview. A renderable hierarchy is maintained while inserting incoming nodes in parallel. This cycle is repeated until the whole hierarchy is constructed.}
  \label{fig:omicron_overview}
\end{figure*}

Rendering a hierarchy while it is under construction is a non-trivial synchronization problem. Since a rendering front can potentially have access to any node in the hierarchy, the use of locks might lead to prohibitive performance. On one hand, using big critical sections by mutexing whole hierarchy levels result in excessive serialization and bad performance. On the other hand, the use of smaller critical sections by mutexing nodes or sibling groups, result in a huge memory overhead to maintain lock data. To efficiently address this problem, one should have a strong definition of what is already processed and is renderable and what is under construction and still volatile.

We propose to synchronize those tasks using specific Morton Curve and Morton Code properties to classify nodes in all curves composing a hierarchy. This classification is based on an Oblique Hierarchy Cut, a novel data-structure to represent hierarchies under construction. Nodes inside an Oblique Cut are guaranteed to be rendered without interference of the construction and vice-versa. An overview of the idea can be seen in Figure~\ref{fig:omicron_overview}. It also shows how new nodes are created and inserted using two operators: \textit{concatenate} and \textit{fix}. Starting from an initial (possibly empty) renderable hierarchy, nodes from the maximum level are inserted using the \textit{concatenate} operator. Then, the hierarchy is evaluated in a bottom-up manner, inserting ancestors of the concatenated nodes into the renderable hierarchy using the \textit{fix} operator.

\begin{figure}[!h]
    \centering
    \includegraphics[width=\columnwidth]{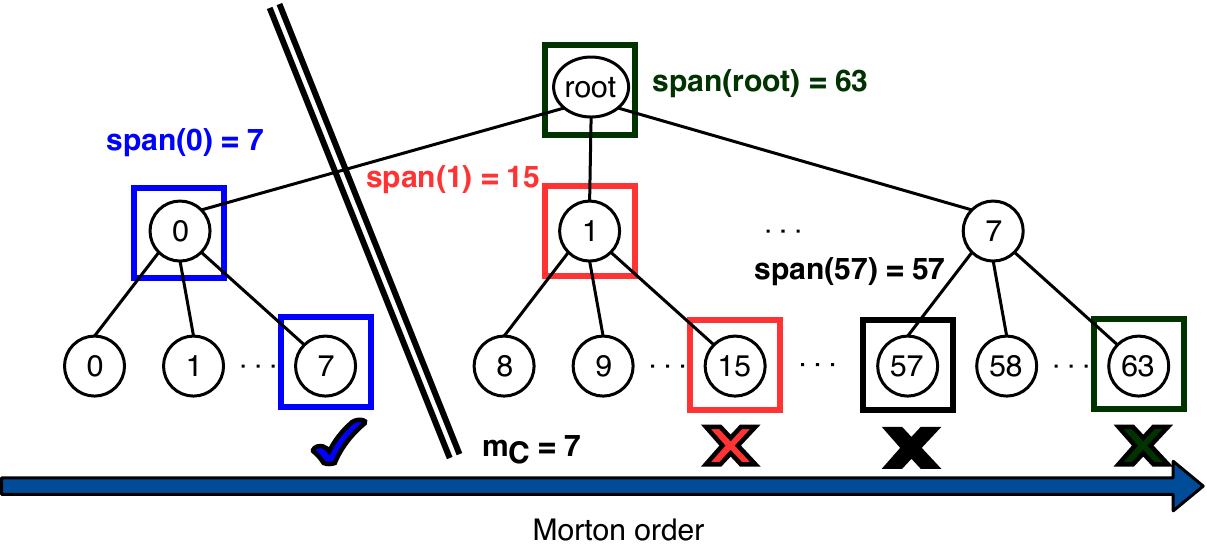}%
    \caption{$span$. In the example, the cut is defined by the delimiting Morton Code $m_C = 7$, defined at the deepest level. Each pair of colored squares shows the input and result of $span$. The blue square case is inside the cut because $span(0) = 7 <= 7$. The other cases (red, green and black) are outside of the cut because $span(x) > 7$. It is important to note that the operation is defined for any level of the hierarchy, even for nodes at the deepest level, where $span(x) = x$.}
    \label{fig:span}
\end{figure}

To evaluate if a node is inside an Oblique Cut we need a methodology that is consistent for all curves at different hierarchy levels. One that makes sense is to consider a node inside the cut if all of its descendants are also inside it. Thus, we need a proper way to relate nodes at Morton Curves at different levels of the hierarchy. For that purpose, let $span(x)$ be a function that returns the Morton Code of the right-most descendant of a supposedly full subtree rooted by $x$. With this definition $span$ has several useful properties. First, it conceptually maps nodes in any hierarchy level with other ones at the deepest level. Thus, it also maps any Morton Curve to the Morton Curve at that level. Not only this, but by definition $span(y) <= span(x)$, for any descendant $y$ of $x$. The cut is then defined as a value $m_C$ at the deepest level and $span$ is used to map any node to its right-most descendant at that level and query if it is left (inside) or right (outside) of the cut. This operation is really efficient because, given the Morton Code of a node, calculating the Morton Code of its right-most descendant is equivalent to concatenating a suffix of bits 1 to that value. Figure~\ref{fig:span} shows how $span$ works.

\section{Oblique Hierarchy Cuts}
\label{section:cuts}

In this section we describe the Oblique Cuts in detail. Given a conceptual expected hierarchy $H$, with depth $l_{max}$, an Oblique Hierarchy Cut $C$ consists of a delimiting Morton code $m_C$ and a set of lists $L_C = \{L_{C,k}, L_{C,k+1} ... L_{C,l_{max}}\}$, where $k$ is the shallowest level of the hierarchy present in the cut. Each node $N$ is uniquely identified by its Morton code $m_N$ and these two concepts are interchangeable from now on. $C$ also has the following important invariants (see Figure~\ref{fig:oblique_cut}):

\begin{figure}[!ht]
    \centering
    \includegraphics[width=0.9\columnwidth]{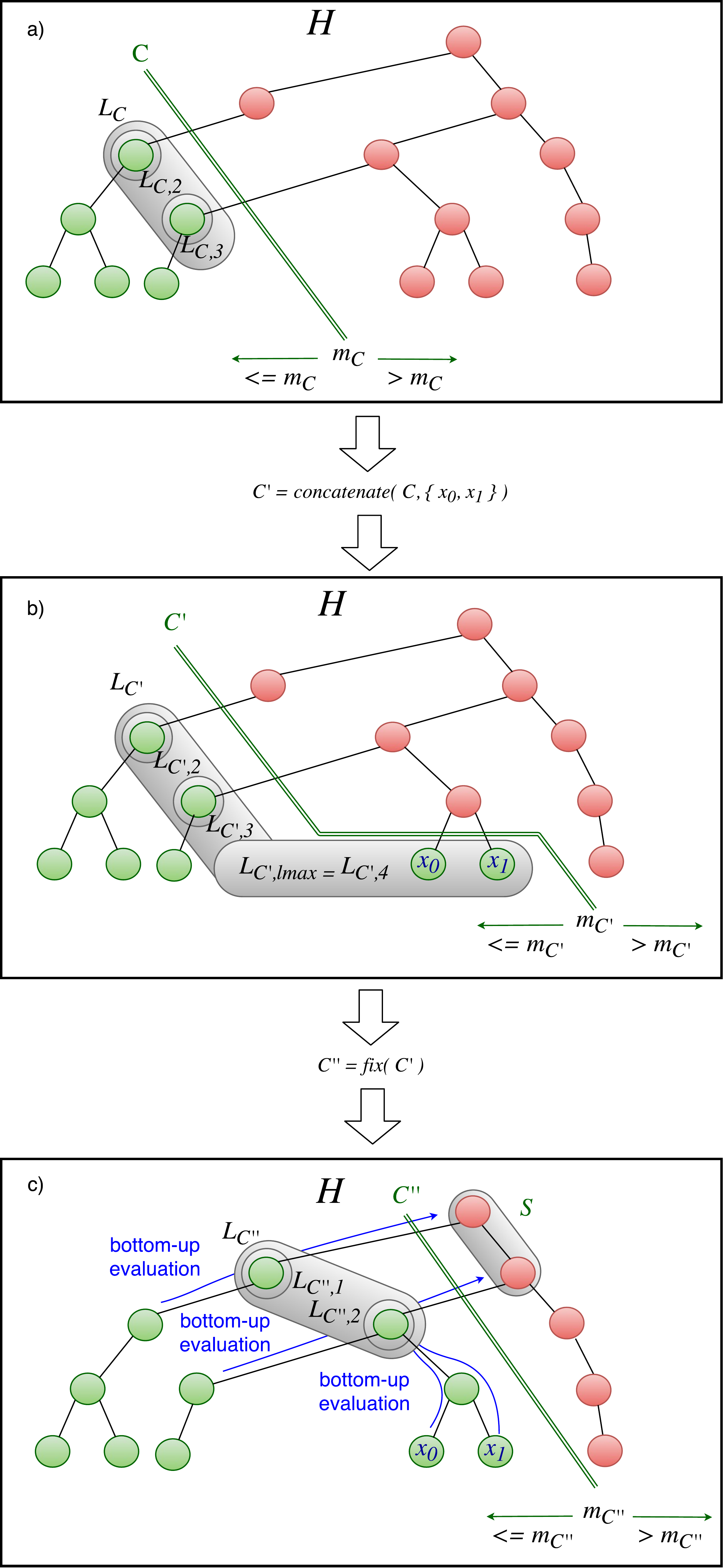}
    \caption{Oblique Hierarchy Cut and operators \textit{concatenate} and \textit{fix}. A cut $C$ is defined by a delimiting morton code $m_C$ and a list of roots per level $L_C$ (a). The green color represent nodes already created and inside the cut. The red color indicates nodes not created yet, which exist only in the conceptual expected hierarchy $H$. The \textit{concatenate} operator inserts new roots $x_0$ and $x_1$ at the deepest level $l_{max}$, resulting in cut $C'$ (b). Then, operator \textit{fix} traverses subtrees bottom-up, creating parents until the boundary $S$ is reached.}
    \label{fig:oblique_cut}  
\end{figure}

\begin{enumerate}[label=1.\arabic*,ref=1.\arabic*]
   \item \label{invariant:morton_code_depth} $m_C$ has level $l_{max}$. 
   \item \label{invariant:list_level} $L_{C,l}$ contains subtrees of $L_{C}$ rooted by nodes at level $l$.
   \item \label{invariant:node_uniqueness} All subtrees in $L_C$ are disjoint.
   \item \label{invariant:morton_order} $L_{C,l}$ is always sorted in Morton order.
   \item \label{invariant:span} All nodes $N$ with $span(m_N) \le m_C$ are in one of the subtrees in $L_C$.
\end{enumerate}

We now formally define the two operators, \textit{concatenate} and \textit{fix}, as well as the important concept of \textit{Placeholder} nodes.

\subsection{Operator \textit{Concatenate}}
\label{sec:concatenate-operator}

The operator \textit{concatenate} is defined as $C' = concatenate(C, \{x_0, ..., x_n\})$ with $m_C < x_0 < ... < x_n$. This operator
incorporates new $l_{max}$ level leaf nodes $\{x_0, ..., x_n\}$ to $C$, resulting in a new cut $C'$. The operator itself is simple and consists of concatenating all new nodes into list $L_{C,l_{max}}$, resulting in $L_{C',l_{max}}$. This operator is illustrated in Figure~\ref{fig:oblique_cut}.

In order for $C'$ to be an Oblique Hierarchy Cut, all invariants must hold. Invariant~\ref{invariant:morton_code_depth} can be maintained by
letting $m_{C'} = x_n$. Invariant~\ref{invariant:list_level} holds by the definition of \textit{concatenate}, since the insertion of the leaf nodes occurs at
the correct list $L_{C,l_{max}}$ at level $l_{max}$. Invariants~\ref{invariant:node_uniqueness} and~\ref{invariant:morton_order} are
ensured by the fact that $m_C < x_0 < ... < x_n$, also established in the definition of \textit{concatenate}. Invariant~\ref{invariant:span}, however,
does not hold, since some of the ancestors $A_x$ of the new nodes $\{x_0, ..., x_n\}$ may have $m_{C} < span(A_x) \le m_{C'}$, but are not in any subtree of $L_{C'}$ after concatenation. In fact, it would be absurd if they were, since all
nodes $N_C$ in $C$ have $span(N_C) \le m_C$ (invariant~\ref{invariant:span}), $m_C < m_{C'}$, and the \textit{concatenate} operator only
inserts nodes greater than $m_C$ at level $l_{max}$.

\subsection{Operator \textit{Fix}}
\label{sec:fix-operator}

To resolve invariant~\ref{invariant:span}, we define the $C'' = fix(C')$ operator, whose purpose is to insert the offending nodes 
in subtrees of $L_{C'}$, resulting in $L_{C''}$, while maintaining all other invariants intact. To achieve this, \textit{fix} first defines the set of offending nodes $A^*_x$ as a subset of $A_x$ with $span(A^*_x) \le m_{C'}$. Second, it identifies all subtree roots in $A^*_x$ whose parents are not in $A^*_x$. Let $S$ be the set of such parent nodes. To identify these subtrees, the lists are processed bottom-up, that is, beginning with $L_{C',l_{max}}$. For each list, its root nodes are visited in Morton order. The evaluation of a list $L_{C',l}$ works in the following manner: identify the sibling root nodes in $L_{C',l}$; check if their parent is in $A^*_x$; create a new subtree rooted by their parents at level $l - 1$; and move the subtrees from $L_{C',l}$ to their respective parent subtrees in $L_{C',l-1}$. Note, however, that if the parent is in $S$ neither the new subtree is created nor its children subtrees are moved. The resulting $L_{C''}$ will have, thus, only subtrees rooted at nodes whose parents are in $S$.

In order to guarantee that \textit{fix} is robust enough, all invariants must be checked for correctness after the operation. Since no new $l_{max}$ level nodes
are inserted by \textit{fix}, we let $m_{C''} = m_{C'}$ and invariant~\ref{invariant:morton_code_depth} is ensured. Invariant~\ref{invariant:list_level}
holds because the $A^*_x$ nodes are inserted in $L_{C'}$ at the same level they are in $H$. Regarding invariant~\ref{invariant:node_uniqueness},
the nodes in $A^*_x$ are unique and they were not in $C'$, since the only nodes $N_{C'}$ that had $span(N_{C'}) > m_C$ were inserted at level $l_{max}$ by the \textit{concatenate} operator. Thus, this invariant holds. Since the subtrees
inserted by \textit{fix} are evaluated in Morton order, they are also inserted in this order, maintaining invariant~\ref{invariant:morton_order}.
Lastly, invariant~\ref{invariant:span} is ensured because the subtrees inserted by \textit{fix} are rooted by nodes whose parents are in $S$, and $S$ is outside of $A^*_x$. Thus, $m_{C''} < span(S)$ and $S$ forms a node boundary outside cut $C''$. The $fix$ operator is illustrated in Figure~\ref{fig:oblique_cut}.

\subsection{Placeholders}
\label{sec:placeholders}


According to the aforementioned definition of Oblique Hierarchy Cut, $H$ can only have leaves at level $l_{max}$, since the \textit{concatenate} operator only inserts nodes at this level. Leaves could be inserted into other levels directly, but it would make it difficult for \textit{fix} to efficiently maintain
invariant~\ref{invariant:morton_order} since the lists $L_{C'}$ are independent and evaluated in a bottom-up manner. To address this issue,
the concept of \textit{placeholder} is defined. A \textit{placeholder} is an empty node at a given level representing a node at a shallower level. More precisely,  
given a node $N$ at level $l$, its placeholder $P_{N,l+1}$ at level $l + 1$ is defined as the rightmost possible child of $N$. In other
words, the Morton code of $P_{N,l+1}$ is $m_N$ followed by a bitmask of as 3 $1$'s as demanded by the
degree of the Octree.  
Note that, with this definition,
$P_{N,l_{max}}$ has Morton code $span(m_N)$, as can be verified by applying the placeholder definition recursively.

A leaf $X$ in $H$ with level
$l < l_{max}$ is represented by placeholder $P_{X,i}$ such that $l < i \le l_{max}$ when inserting the subtree of level $i$ at $L_{C'_i}$.
Placeholders are used as roots of degenerate subtrees, since there is no purpose for them inside subtrees. Even if not meaningful for $H$,
placeholders ensure invariant \ref{invariant:morton_order} in \textit{fix} until level $l$ is reached. Figure \ref{fig:cut_progression} shows the concept of placeholders. 

\subsection{Sequence of Oblique Hierarchy Cuts}
\label{sec:sequence-cuts}

\begin{figure}[!ht]
	\centering
	\includegraphics[width=\columnwidth]{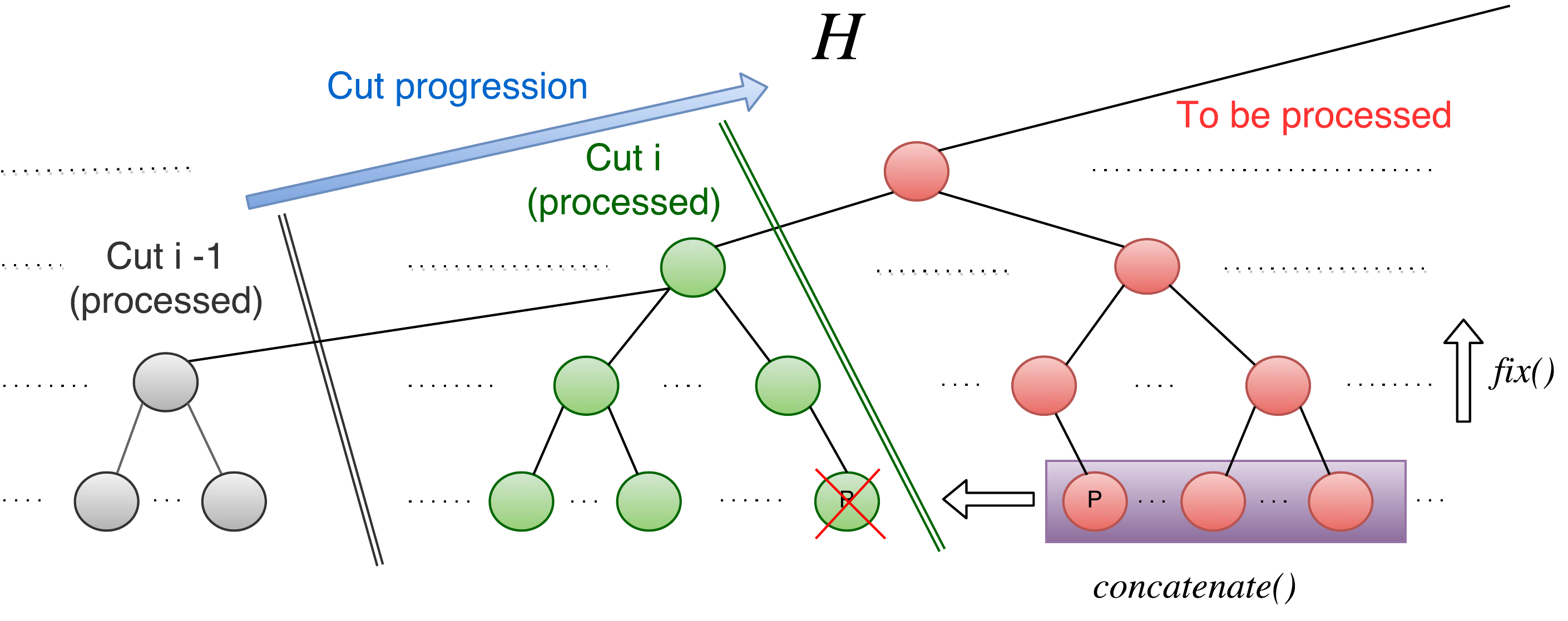}
	\caption{Oblique Hierarchy Cut progression. As operators \textit{concatenate} and \textit{fix} are used, the cuts sweep their associated hierarchy $H$.
  	Placeholders are marked with a P and the ones used but removed while processing lists bottom-up are also marked with a red X.}
	\label{fig:cut_progression}
\end{figure}

Intuitively, a sequence of Oblique Hierarchy Cuts $C_i$ resulting from sequentially applying operators \textit{concatenate} and \textit{fix} until no
more leaf nodes or placeholders are left for insertion results in an oblique sweep of $H$, as can be seen in Figure~\ref{fig:cut_progression}. To prove this, let $C_{end}$ be the last cut in this sequence. Because of invariant~\ref{invariant:span}, all
nodes $N$ in $H$ with $m_N \le m_{C_{end}}$ will be in subtrees in $L_{C_{end}}$ after $fix$. Since there are no more placeholders or leaf
nodes in level $l_{max}$, there are no nodes $N$ with $m_N > m_{C_{end}}$ and, thus, $S$ is composed only by the $null$ node (parent of $H$'s root node). Since there are
no other parents outside the subtrees that have roots with parents in $S$, and $S$ has only a single element, $L_{C_{end}}$ is composed by a
single subtree, named $T$. Also, $T$'s root has parent equal to the $null$ node. Thus, $T = H$, as intuitively suspected.

\section{Oblique Hierarchy Cut Front}
\label{section:front}

\begin{figure}[htb]
	\centering
	\includegraphics[width=0.9\columnwidth]{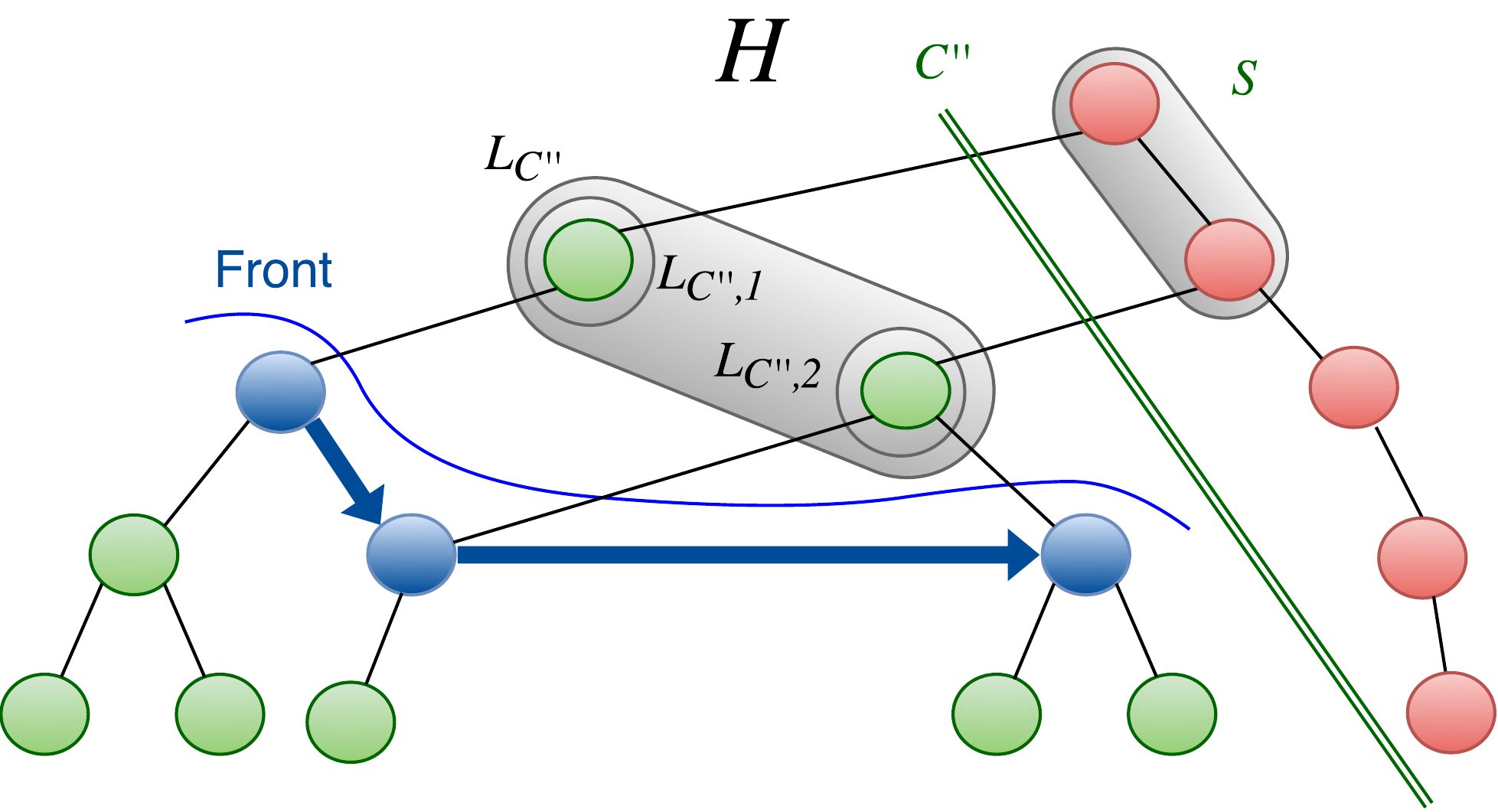}
	\caption{Example of valid Oblique Hierarchy Cut Front. The direction of the blue arrows indicate the order restriction imposed by
		invariant~\ref{invariant:front_order}. The fact that all nodes in the front are not roots in $L_C$ ensures invariant~\ref{invariant:front_roots}.}
	\label{fig:front}
\end{figure}

Concomitantly with the building of $H$ with progressive oblique cuts, a rendering process might be traversing
the already processed portions of $H$ with the help of a front (see Figures~\ref{fig:omicron_overview} and \ref{fig:front}). Thus, for a 
given Oblique Hierarchy Cut $C$, the rendering process will adaptively maintain a front $F_C$ restricted to 
the renderable part of $H$. 
In order to ensure proper independence of $F_C$ with respect to $C$ and other
important properties needed later, we define two invariants:

\begin{enumerate}[label=2.\arabic*,ref=2.\arabic*]
 \item \label{invariant:front_order} If $F_C$ is composed of $n$ nodes, named $F_{C,i}$, with $1 \le i \le n$, then
 $span( F_{C,1} ) < ... < span( F_{C,i} ) < span( F_{C,{i+1}} ) < ... < span( F_{C,n} )$.
 \item \label{invariant:front_roots} 
The roots of subtrees in $L_C$ cannot enter the Front.
\end{enumerate}

Invariant~\ref{invariant:front_order} ensures that sibling nodes will be adjacent in the front, which eliminates searches and simplifies the
\textit{prune} operation. Invariant~\ref{invariant:front_roots} is defined because the roots of subtrees in $L_C$ 
are being moved among lists by the \textit{fix} operator in order to create subtrees at other levels and thus are not safe to enter the front.
Note that both invariants impose restrictions on the \textit{prune} operator in order to ensure that all nodes on the front 
are roots of disjoint subtrees and do not include nodes still being processed.
Similarly, placeholders cannot be pruned either since their parents might not yet be defined.


In summary, the evaluation of an Oblique Hierarchy Cut Front consists of three steps:
\begin{enumerate}
	\item Concatenate new placeholders into the front.
	\item Choose the hierarchy level $l$ where candidates for substituting placeholders in the front are to be sought.
	\item Iterate over all front nodes, testing whether they are placeholders that can be substituted, and whether they need to be pruned, branched or rendered.
\end{enumerate}

Leaf insertions and placeholder substitutions will be further described in the next sections. The other aspects of operators \textit{prune} and \textit{branch} work as usual. All valid inner nodes are reachable by \textit{prune} operations from the leaves, ensuring proper rendering capabilities for the cut. An example of a valid Oblique Hierarchy Cut Front is given in Figure~\ref{fig:front}.

\subsection{Insertion of new nodes}
\label{sec:front-insertion}

Since the root of $H$ is only available after all sequential cuts are evaluated, the usual front initialization is not possible for $F_C$. To insert nodes in the Oblique Hierarchy Cut Front two operators are used: \textit{insertPlaceholder} and \textit{insertLeaf}. In order to simplify leaf and placeholder insertion and substitution, all leaves are first inserted in the front as placeholders and saved in a per-level list of leaves to be replaced. One main reason for this duplication is that new nodes are always inserted as roots in $L_{C, l_{max}}$, and cannot be in the front due of invariant~\ref{invariant:front_roots}. Thus, placeholders mark their position until the \textit{fix} operator moves them to other subtrees. The front is, then, continuously checked to see if placeholders can be replaced by leaf nodes. This substitution is detailed in the next section.

The \textit{insertPlaceholder} operator in its turn is simple since it can just concatenate placeholders at the end of the front. This maintains the invariants since placeholders are available at level $l_{max}$ and they are processed in Morton order by \textit{fix}.

\subsection{Substitution of placeholders}
\label{sec:front-substitution}

Since the leaf lists are organized by level, and the placeholders and leaves are respectively inserted into the front and into the lists in Morton order, a very simple and
efficient substitution scheme is proposed. Given a placeholder and a substitution level $l$, it consists in verifying if the first element in the
leaf list of level $l$ is an ancestor of the placeholder. If it is, the leaf is removed from the substitution list and replaces the
placeholder in the front. Since comparison of Morton codes
is a fast $O(1)$ operation, the entire placeholder substitution algorithm
is also $O(1)$.

Keeping in mind that for each front evaluation a single level $l$ will be checked for substitution, all leaves at level $l$ are guaranteed to be substituted in a single front evaluation. To verify this, note that if $P_{i}$ and $P_{i + 1}$ are sequential placeholders at the same level and $L_j$ and $L_k$
are their leaf substitutes, then $k = j + 1$. This comes again from the fact that all insertion lists and front nodes at a given level are in Morton order 
and that a leaf and its placeholder have a one-to-one relationship. Thus, if $P_{i}$ is substituted and, as a consequence $L_j$ is removed from the
substitution list, then the new first leaf in that list will be $L_k$, resulting in $P_{i+1}$ being the next placeholder to be successfully
substituted at that level. Consequently, for each placeholder in the front we need only to verify the first leaf of the list, and after one evaluation the list for level $l$ will be emptied.

\subsection{Choice of substitution level}
\label{sec:front-substitution-level}

In order to maximize node substitution, $l$ is chosen as the level with most insertions. This is an obvious choice, since the list will be completely emptied after the evaluation, so we are substituting the maximum number of placeholders in one iteration. The nodes not substituted in the current front evaluation
are ignored since their corresponding leaves are not in level $l$. However, the algorithm guarantees that all currently inserted leaves will substitute their placeholders in the next $l_{max} - 1$ front evaluations at max. Thus, the delay to starting rendering a leaf node after insertion is minimal.

\section{Sample OMiCroN implementation}
\label{section:omicron}

We have developed a multi-threaded implementation of the OMiCroN algorithm in C++ where the splat rendering is done on GPU. The implementation follows the algorithms outlined in the previous sections, but a few adaptations are necessary with regard to concurrency control.





A sorted input stream feeds a master thread that organizes worklists for the current level $l$. They consist of nodes for the fix operator, which are distributed among the master and slaves threads for processing. 
To simplify distribution, the worklists have fixed size. As a consequence, a sibling group can be split between threads, which might lead to parent node duplication. Thus, after a processing iteration, the master thread checks the first and last nodes of the resulting adjacent worklists at level $l-1$ and eliminates duplicates, moving the children of the eliminated nodes accordingly, so no child is lost.
This movement of children resulting from node duplication imposes an additional restriction to the nodes that can be added to the front. Therefore, invariant~\ref{invariant:front_roots} defined in Section~\ref{section:front} must be modified
so no volatile nodes enter the front, becoming:
\begin{enumerate}
 \item[2.2] \label{invariant:front_roots_parallel} The roots of subtrees in $L_C$ \textbf{and their children} cannot enter the front.
\end{enumerate}

Since worklist sizes are expected to become smaller and smaller as the hierarchy is traversed bottom-up, the master thread also applies simple load balancing heuristics by merging worklists as they are tested for duplicates. 
Once level $l$ is processed, OMiCroN checks the amount of work available at level $l-1$. More precisely, it compares the available work at level $l-1$ with level $l_{max}$ to verify if it is worth continuing the current $fix$ pass, or if it is better to start another $fix$ pass from scratch. 

In order to maintain the use of main memory within a given budget, it is also possible to enable
a very simple optimization, called \textit{Leaf Collapse}. This optimization removes all leaves at level $l_{max}$ which form
a chain structure with their parents, i.e., leaves that do not have siblings.

Rendering itself is performed in a separate front tracking thread. This thread is signaled the availability of newly processed data by the master thread, thus requiring synchronization. This drawback is minimized by having a different lock per hierarchy level. Another efficiency tweak consists of segmenting the front evaluation along several frames in order to amortize its cost.
A simple rendering approach based on \emph{splats}~\cite{Rusinkiewicz2000} is used in our experiments. OMiCroN nodes contain point splats defined by a center point and two tangent vectors $u$ and $v$. Parent node creation follows a policy that tries to maintain the ratio between the number of points in a parent and its children, where a parent contains a subset of the splats in its children with scaled tangent vectors.

For each frame, the front or front segment is evaluated based on the projection threshold. If the projection of a given node is sufficiently large in comparison to the threshold, it suffers branching and its children sibling group is pushed into the rendering queue. Conversely, if it is sufficiently small, it suffers pruning and its parent is pushed into the queue. Otherwise, the node stays in the front to be rendered. The rules for branching and pruning are the ones discussed in Section~\ref{section:front}. Finally, the splats in the rendering queue are used as input for the traditional two-pass EWA filter described in \cite{Zwicker2001}. Several methods for computing the sizes of the projected splats were tested \cite{Zwicker2001,Zwicker2004,Botsch2004,Weyrich2007}. The splat bounding box computation algorithm described in~\cite{Weyrich2007} resulted in the best performance-quality relationship and all results reported in this paper applied it.

\section{Experiments}
\label{section:comparisons}

The prototype implementation was tested using four point cloud datasets obtained at the \href{https://graphics.stanford.edu/projects/mich/}{Digital Michelangelo Project page}: David (469M points, 11.2GB), Atlas (255M points, 6.1GB), St. Mathew (187M points, 4.5GB) and Duomo (100M points, 2.4GB). The maximum hierarchy depth was set to 7 to ensure memory footprints compatible with available memory and swap area. Coordinates in all datasets were normalized to range $[0,1]$.

\subsection{Rendering latency tests}
\label{section:rendering_tests}

In order to assess the actual delay from the moment the raw unsorted collection of points is available to the moment where rendering actually starts, we must consider the sorting process at some length. The simplest scenario consists of a separate thread that reads the whole collection, sorts it and streams it to OMiCroN. In this case, OMiCroN must wait at least for the whole collection to be read by the sorting application, and for the sort itself. In a more elaborate setup, the sorting process might start feeding OMiCroN as soon as a prefix of the sorted collection becomes available. These two scenarion are variations of pipeline 1 in Fig.~\ref{fig:pipelines}. In order to measure these gains, we conducted a set of experiments. Our testbed consists of a desktop computer with an Intel Core i7-3820 processor with 16GB memory, NVidia GeForce GTX 750 and a SanDisk 120GB SSD. The same SSD is used for both swap and I/O.

\begin{figure*}[!ht]
\centering
\subfloat[St. Mathew.]{
    \includegraphics[width=.32\textwidth]{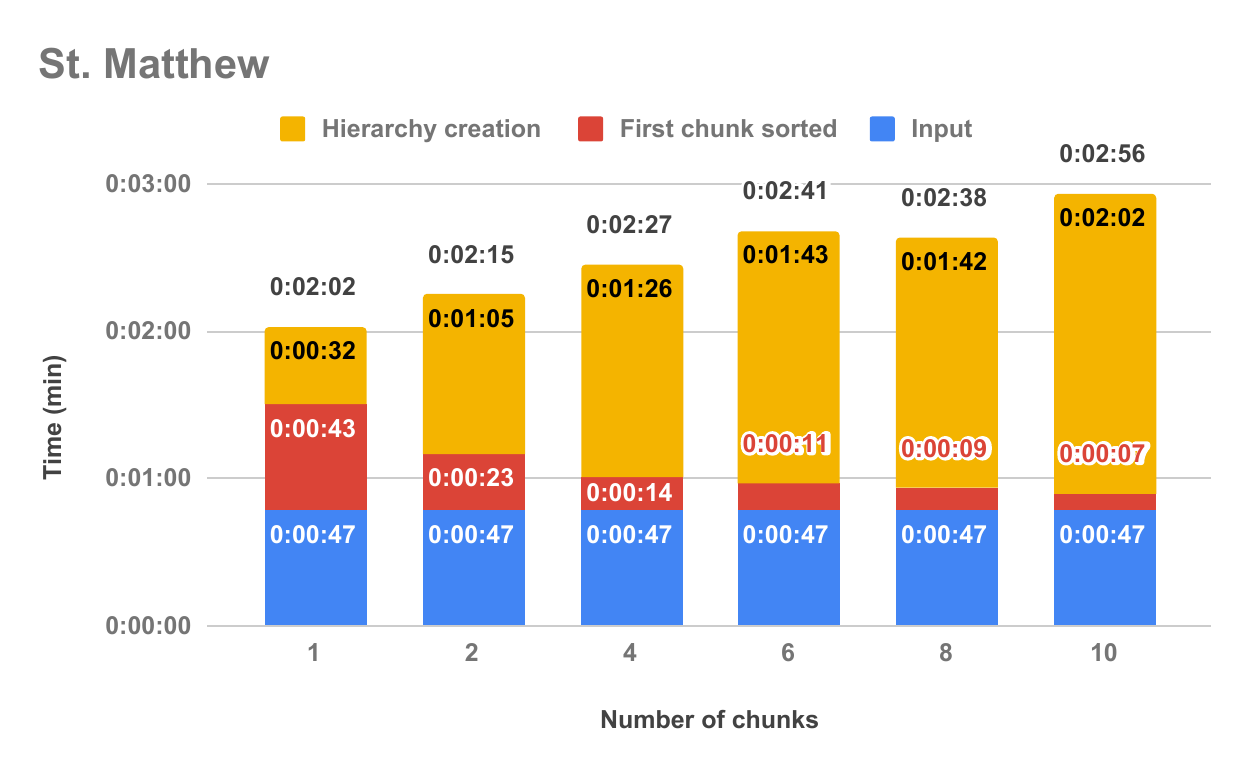}%
    \label{fig:partial_sort_mathew}
}
\hfil
\subfloat[Atlas]{
  \includegraphics[width=.32\textwidth]{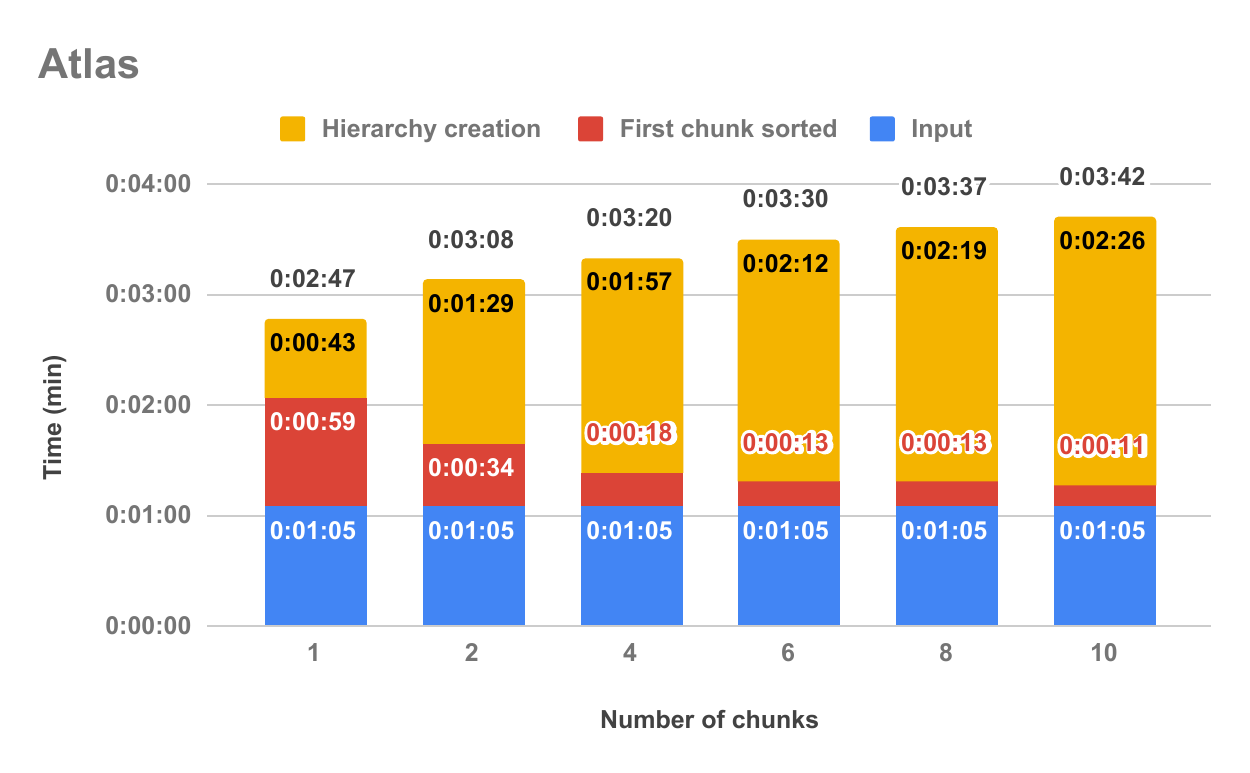}%
  \label{fig:partial_sort_atlas}
}
\hfil
\subfloat[David]{
  \includegraphics[width=.32\textwidth]{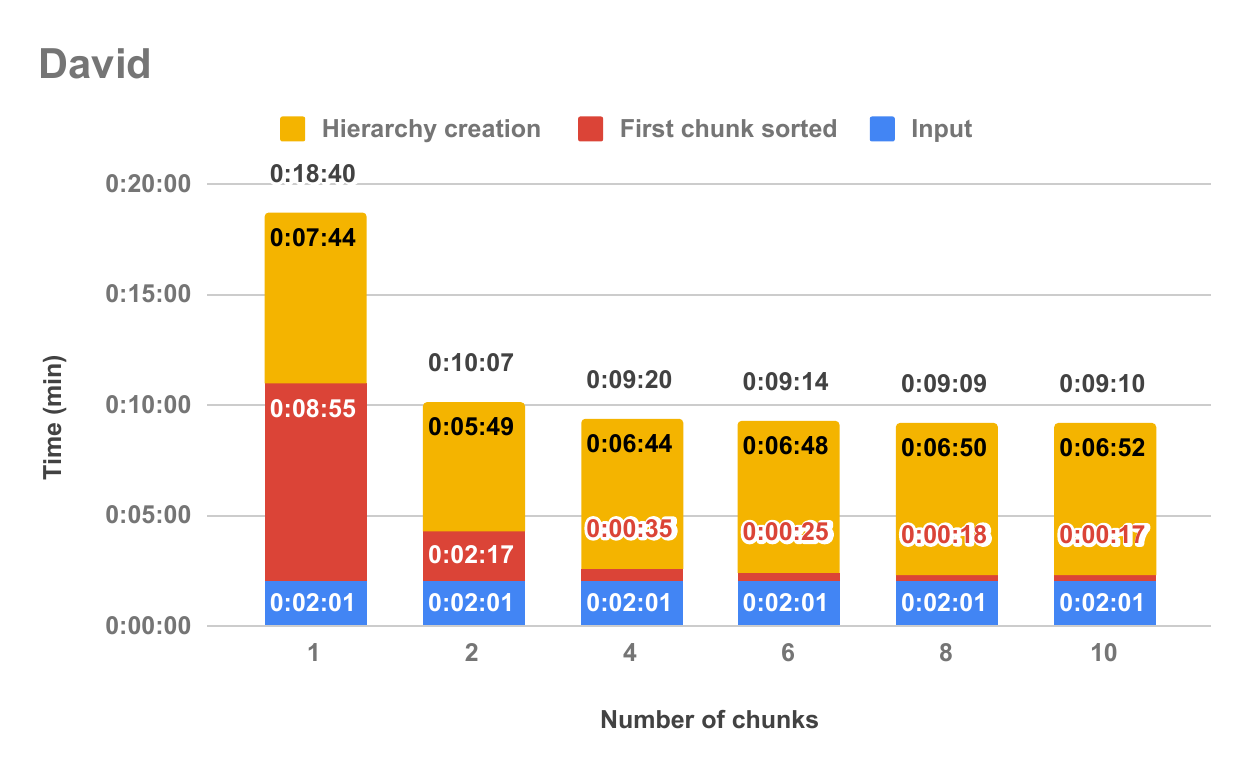}%
  \label{fig:partial_sort_david}
}
\caption{Impact of the number of sort chunks. After a constant amount of time spent reading the input (blue), the first chunk is sorted (red), starting the parallel hierarchy creation and rendering (orange). The first column in all charts corresponds to the case where all input is sorted before the hierarchy creation begins. }
\label{fig:partial_sort}
\end{figure*}

The first experiment consists of consecutively sorting and streaming chunks of the input to OMiCroN. We use the parallel IntroSort available in the Standard Template Library (STL) of the C++ programming language (std::partial\_sort() or std::sort()). Parallel rendering and leaf collapse are enabled for these tests. Since rendering starts as soon as the first sorted chunk becomes available, using more chunks allows rendering to start earlier, as shown in Figure~\ref{fig:partial_sort}. In particular, increasing the number of sorting chunks can improve the time between the moments input finishes and rendering starts from 5 to 31 times, depending on the size of the dataset. The price of this early rendering is that hierarchy creation time may increase up to 4 times, also depending on the dataset size. For large datasets, the partial sort can diminish the use of swap during sort and hierarchy creation, resulting in better timings in all aspects, as Figure~\ref{fig:partial_sort_david} demonstrates. We also noted that OMiCroN consumes sorted chunks almost as fast as they are produced and streamed, and the hierarchy is finished at most $1s$ after the last byte of the sorted stream is read. Another conclusion is that the number of chunks represents a trade-off between the time for starting rendering and the total time to sort the dataset. The exception for this rule is the David dataset.




\begin{figure}[!ht]
  \centering
  \includegraphics[width=\columnwidth]{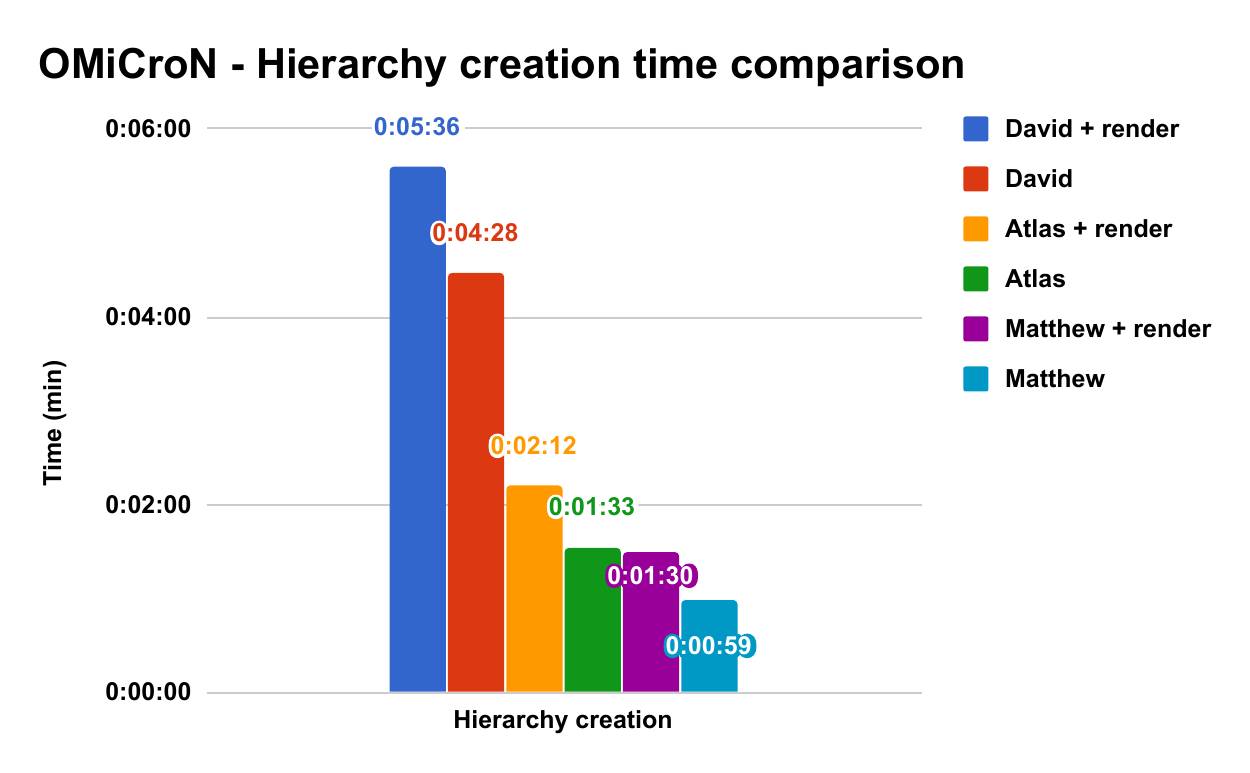}%
  \caption{Comparison of hierarchy creation with and without parallel rendering. Sorted data is streamed directly from disk. The overhead imposed by parallel rendering is between 20\% (David) and 34\% (St. Mathew). 
  }
  \label{fig:hierarchy_creation}
\end{figure}

The second experiment consists of profiling and comparing OMiCroN with the parallel rendering activated and deactivated at hierarchy creation time, also evaluating the system core usage while running the algorithm. The purpose of this test is to measure the overhead of parallel rendering and the overall usage of resources. The input for this test consists of datasets sorted in Morton order and the data is streamed directly from disk (pipeline 2 in Fig.~\ref{fig:pipelines}). Leaf collapse is disabled. Figure~\ref{fig:hierarchy_creation} shows the results.
The overhead imposed is between 20\% (David) and 34\% (St.Mathew), which is an evidence that the overhead impact decreases as the dataset size increases. This is a desirable property for an algorithm designed to handle large datasets. The final observation from this experiment is that OMiCroN maintains the usage of all 8 logical cores near 90\% with peaks of 100\% for the entire hierarchy creation procedure, with parallel rendering enabled or disabled. This fact justifies OMiCroN's fast hierarchy creation times.

\begin{figure}[!t]
  \centering
  \includegraphics[width=\columnwidth]{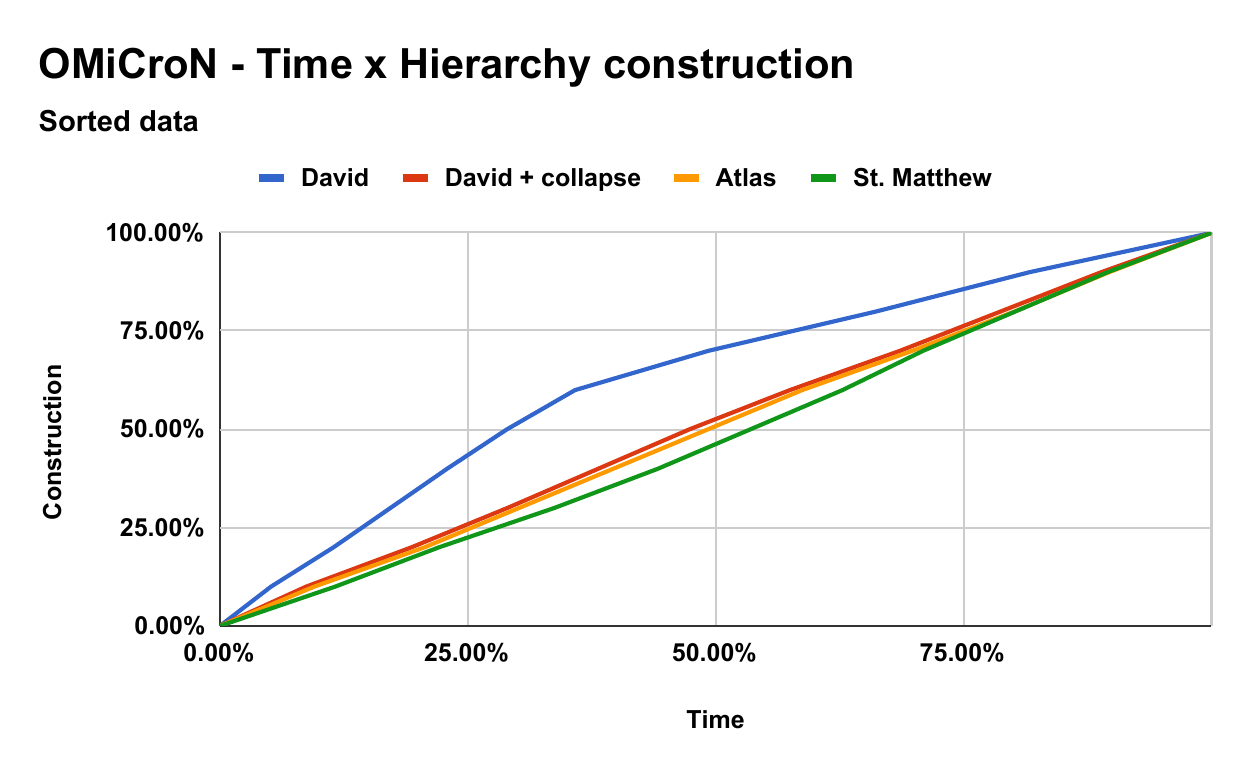}%
  \caption{Hierarchy creation over time. Sorted data is streamed directly from disk, parallel rendering is enabled and leaf collapse is disabled unless pointed otherwise. 
  }
  \label{fig:hierarchy_over_time}
\end{figure}

The third experiment's purpose is to generate data for better understanding the hierarchy creation progression over time. It consists of measuring the time needed to achieve percentile milestones of hierarchy creation. The best scenario is a linear progression over time so new data can be presented smoothly to the user while the hierarchy is being constructed. For this test, the sorted data is streamed directly from disk, parallel rendering is enabled and leaf collapse is disabled unless noted otherwise. The results are presented in Figure~\ref{fig:hierarchy_over_time}. We can conclude that the hierarchy construction has the expected linear progression. The exception is the David dataset with leaf collapse disabled. This behavior is caused by the hierarchy size, which exceeds available memory, forcing the use of swap area and performance degradation. When leaf collapse is enabled, swap is avoided and the behaviour is again linear, as Figure~\ref{fig:hierarchy_over_time} also demonstrates.

\subsection{Hierarchy creation and rendering}




%

\begin{figure*}[!t]
\centering
\subfloat[David, leaf collapse on, 0.2 point ratio.]{
  \includegraphics[width=.3\textwidth]{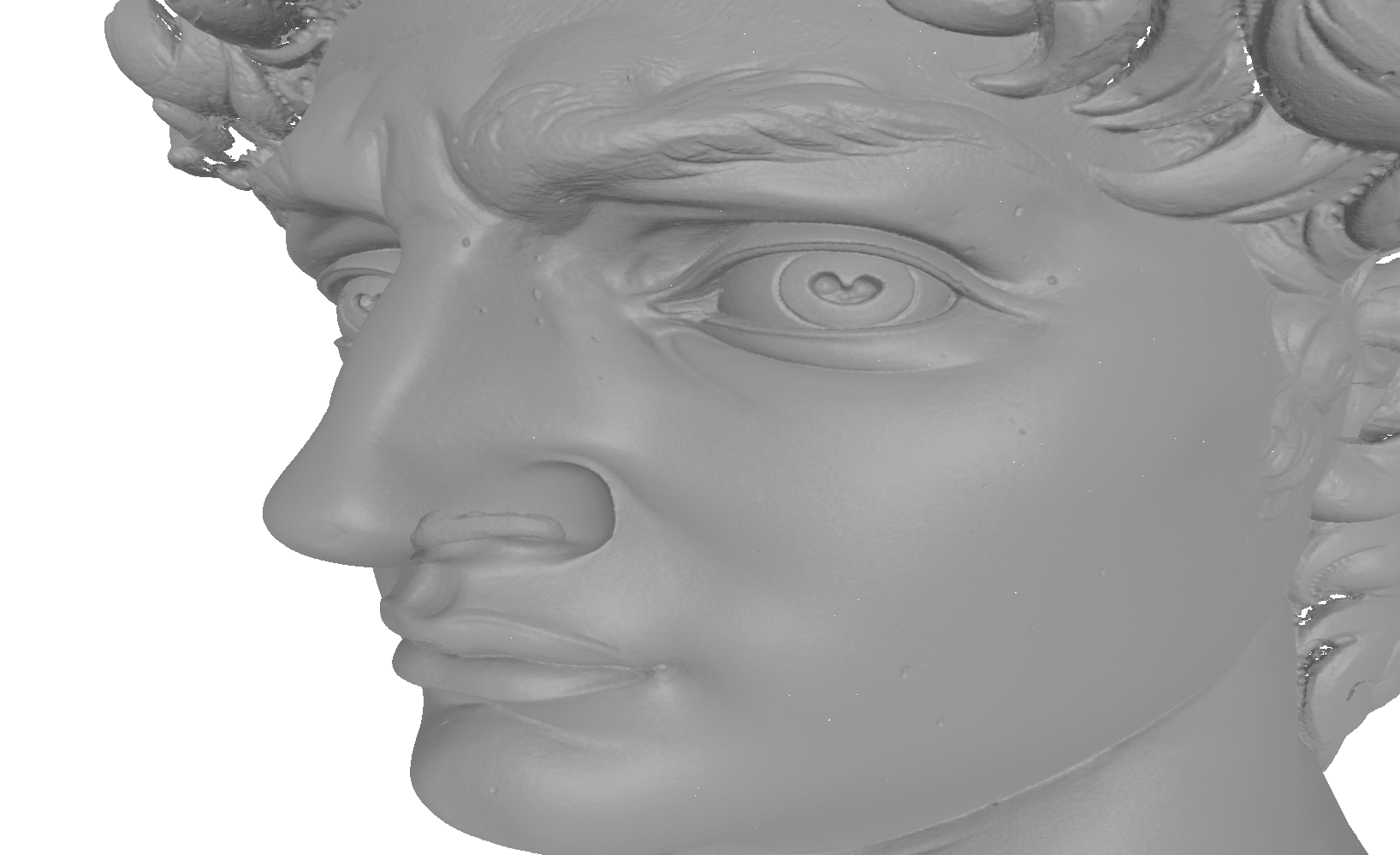}%
  \label{fig:david_leaf_collapse_02_parent_ratio}
}
\hfil
\subfloat[David, leaf collapse on, 0.25 point ratio.]{
  \includegraphics[width=.3\textwidth]{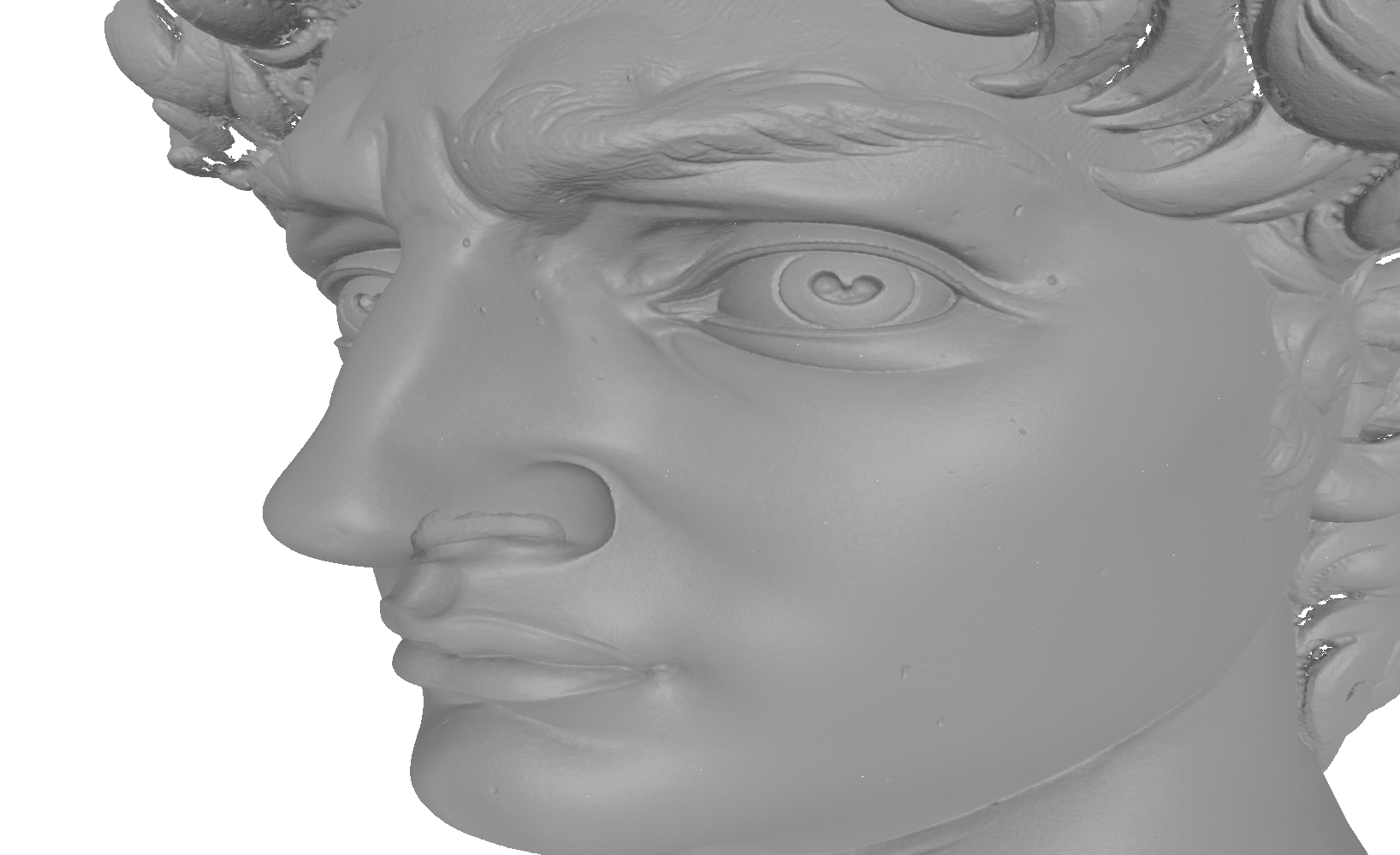}%
  \label{fig:david_leaf_collapse_025_parent_ratio}
}
\hfil
\subfloat[David, leaf collapse off, 0.25 point ratio.]{
  \includegraphics[width=.3\textwidth]{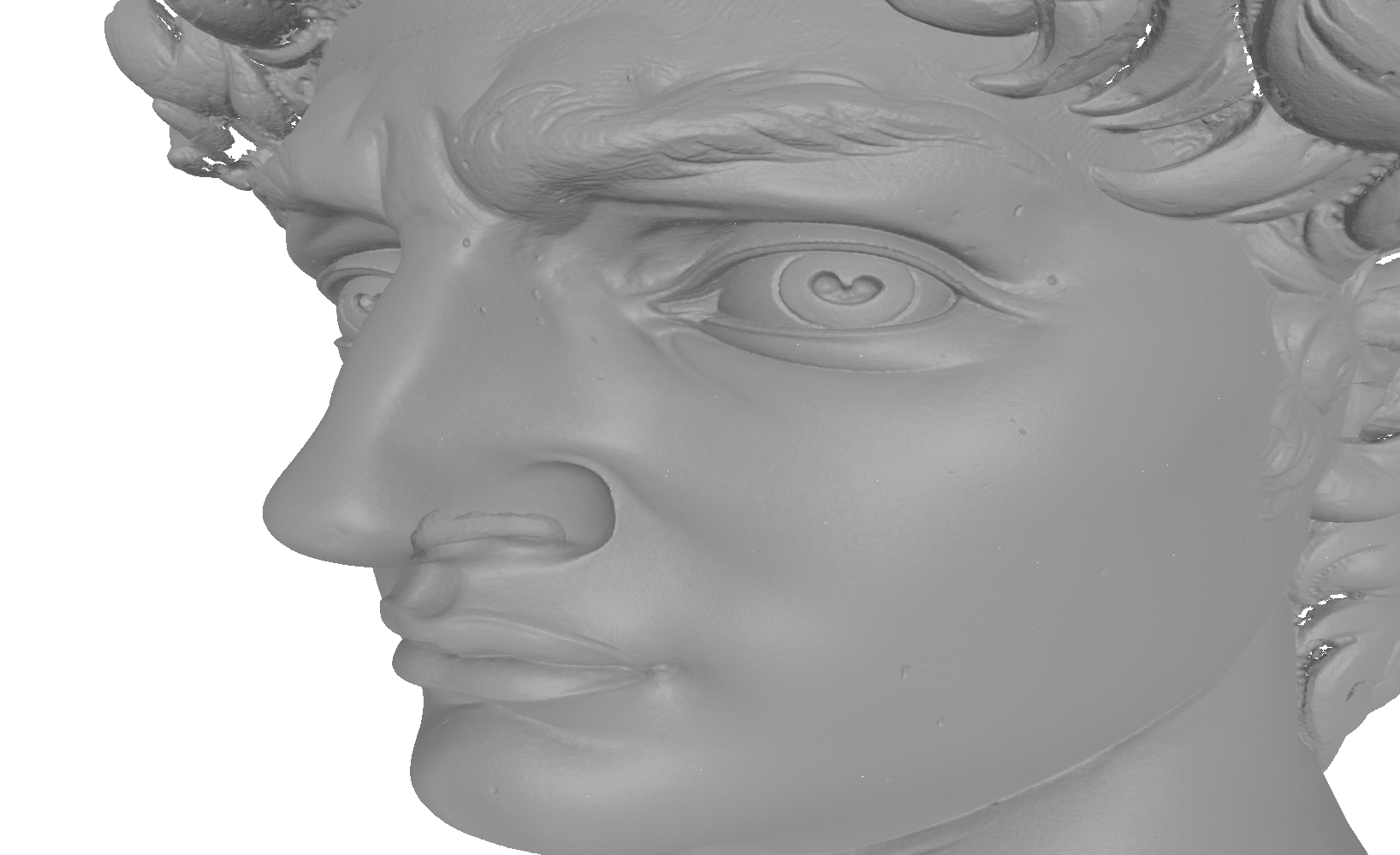}%
  \label{fig:david_no_leaf_collapse_025_parent_ratio}
}
\hfil
\subfloat[Atlas, leaf collapse on, 0.2 point ratio.]{
  \includegraphics[width=.3\textwidth]{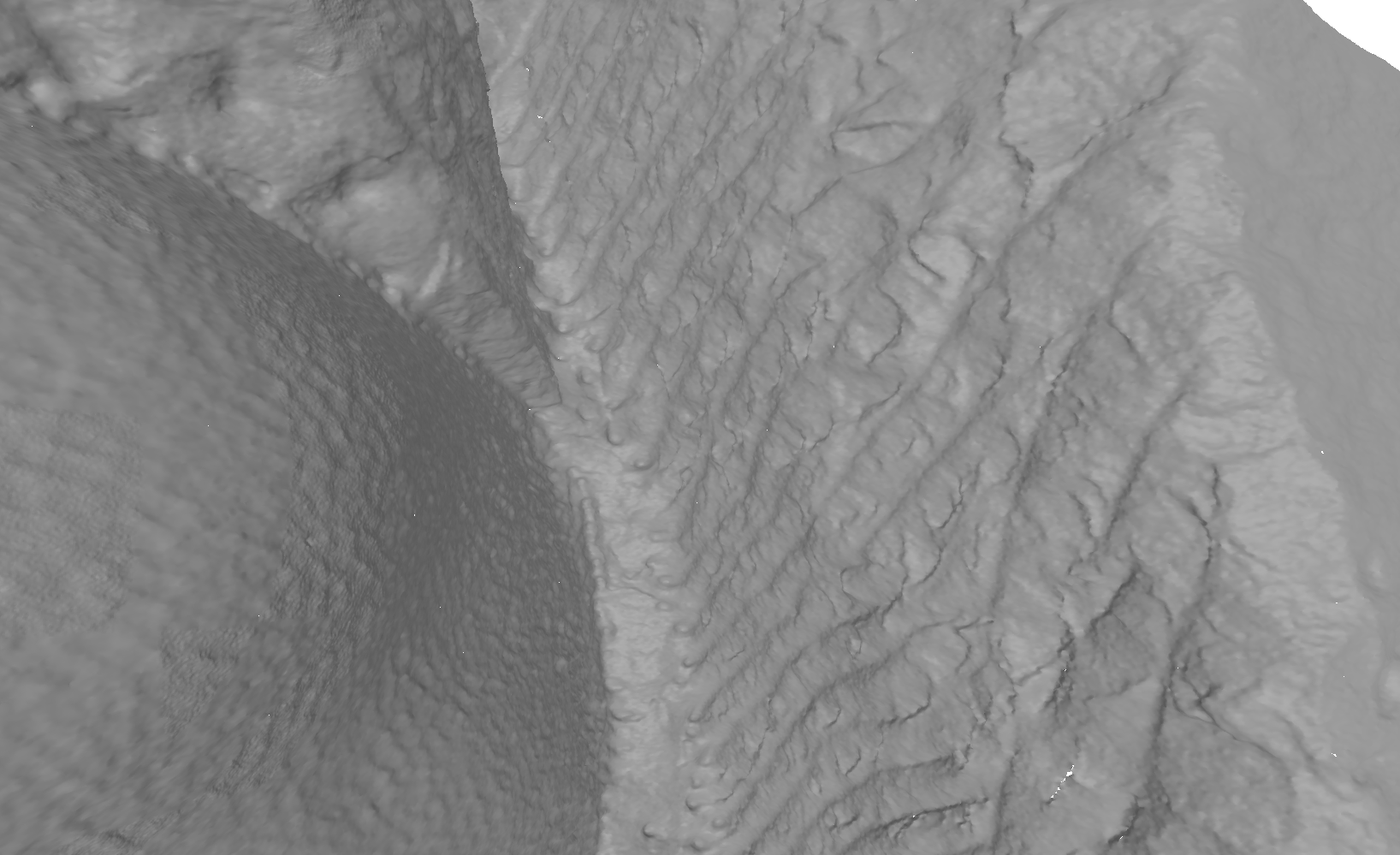}%
  \label{fig:atlas_leaf_collapse_02_parent_ratio}
}
\hfil
\subfloat[Atlas, leaf collapse on, 0.25  point ratio.]{
  \includegraphics[width=.3\textwidth]{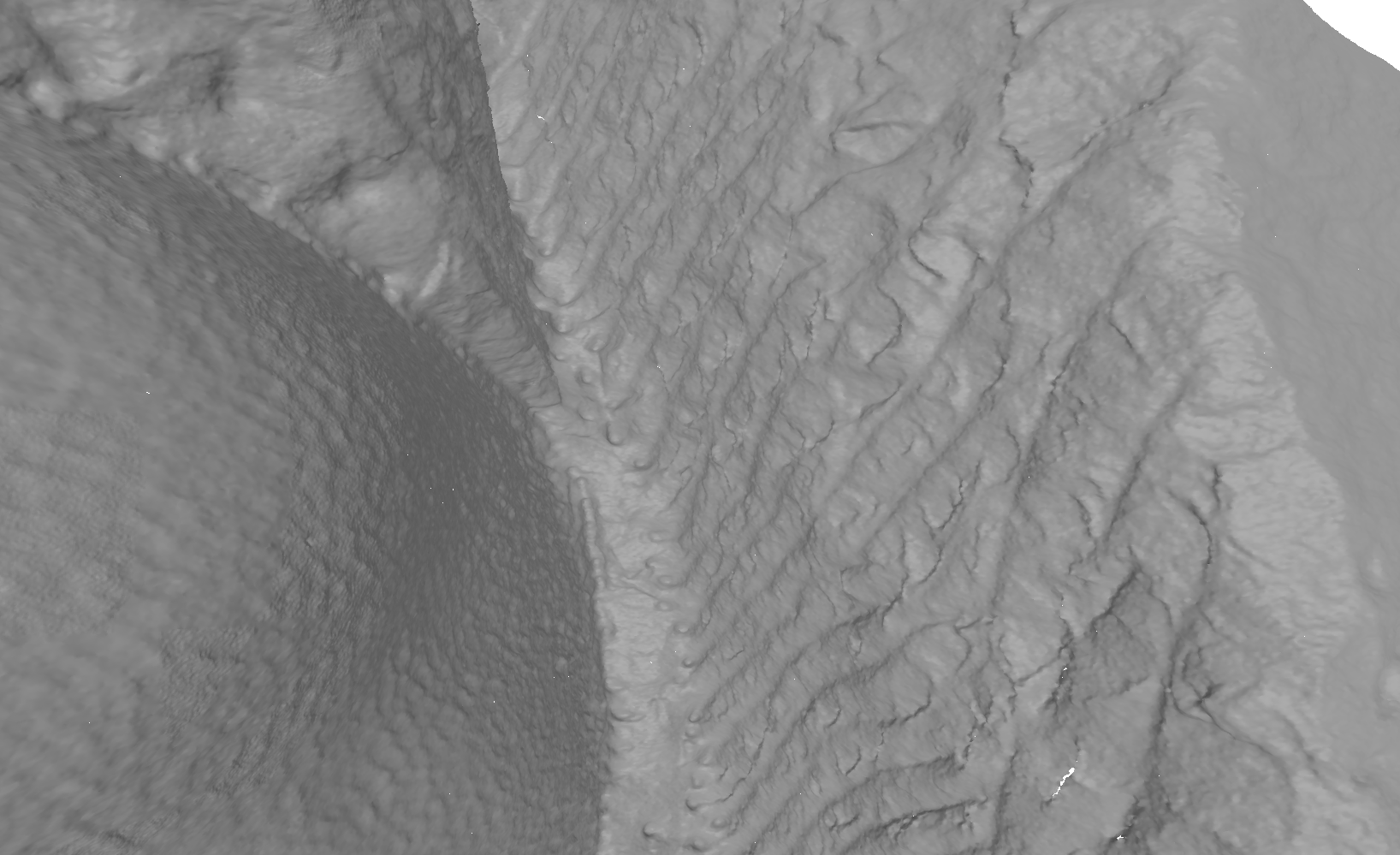}%
  \label{fig:atlas_leaf_collapse_025_parent_ratio}
}
\hfil
\subfloat[Atlas, leaf collapse off, 0.25 point ratio.]{
  \includegraphics[width=.3\textwidth]{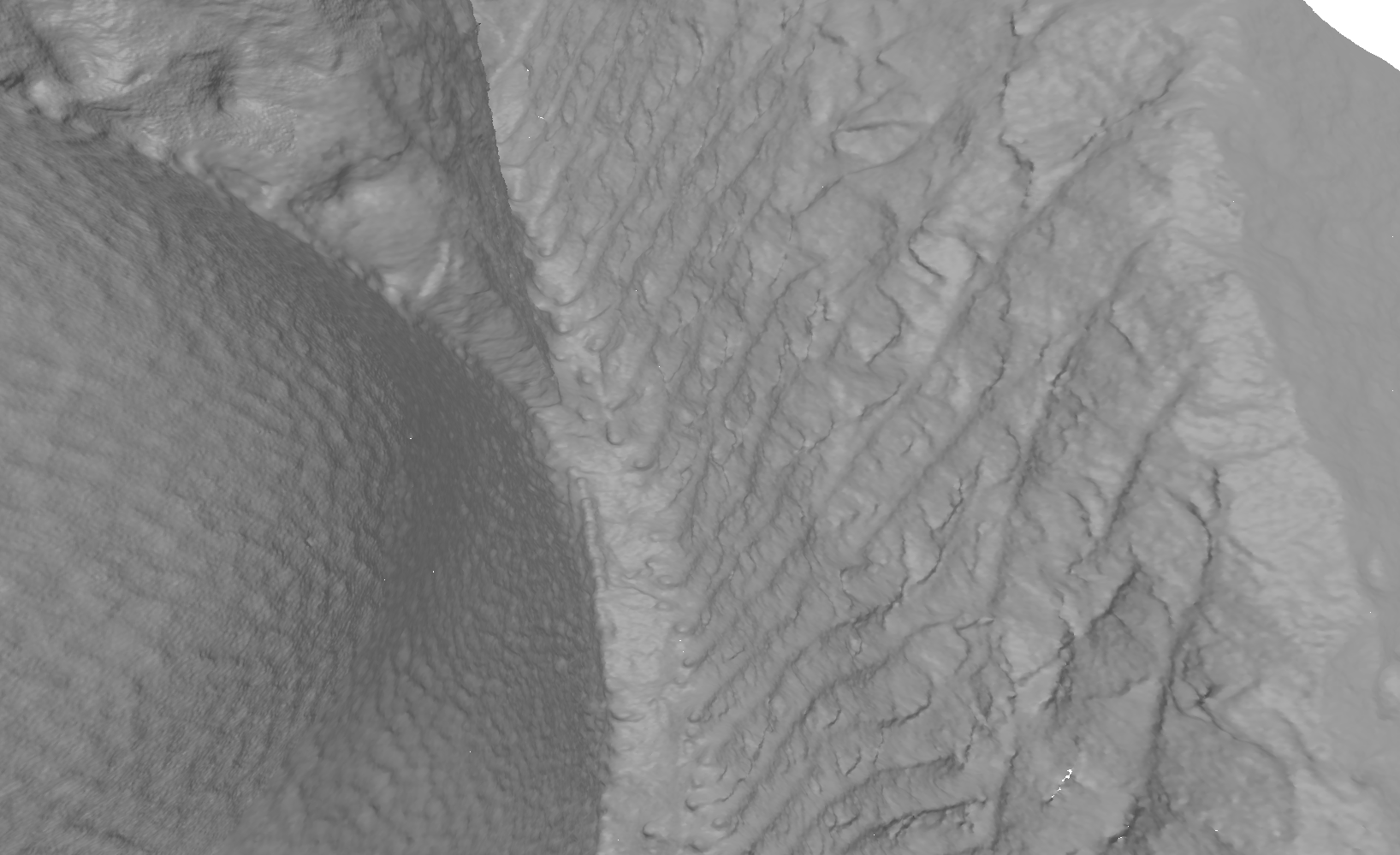}%
  \label{fig:atlas_no_leaf_collapse_025_parent_ratio}
}
\hfil
\subfloat[St. Mathew, leaf collapse on, 0.2 point ratio.]{
  \includegraphics[width=.3\textwidth]{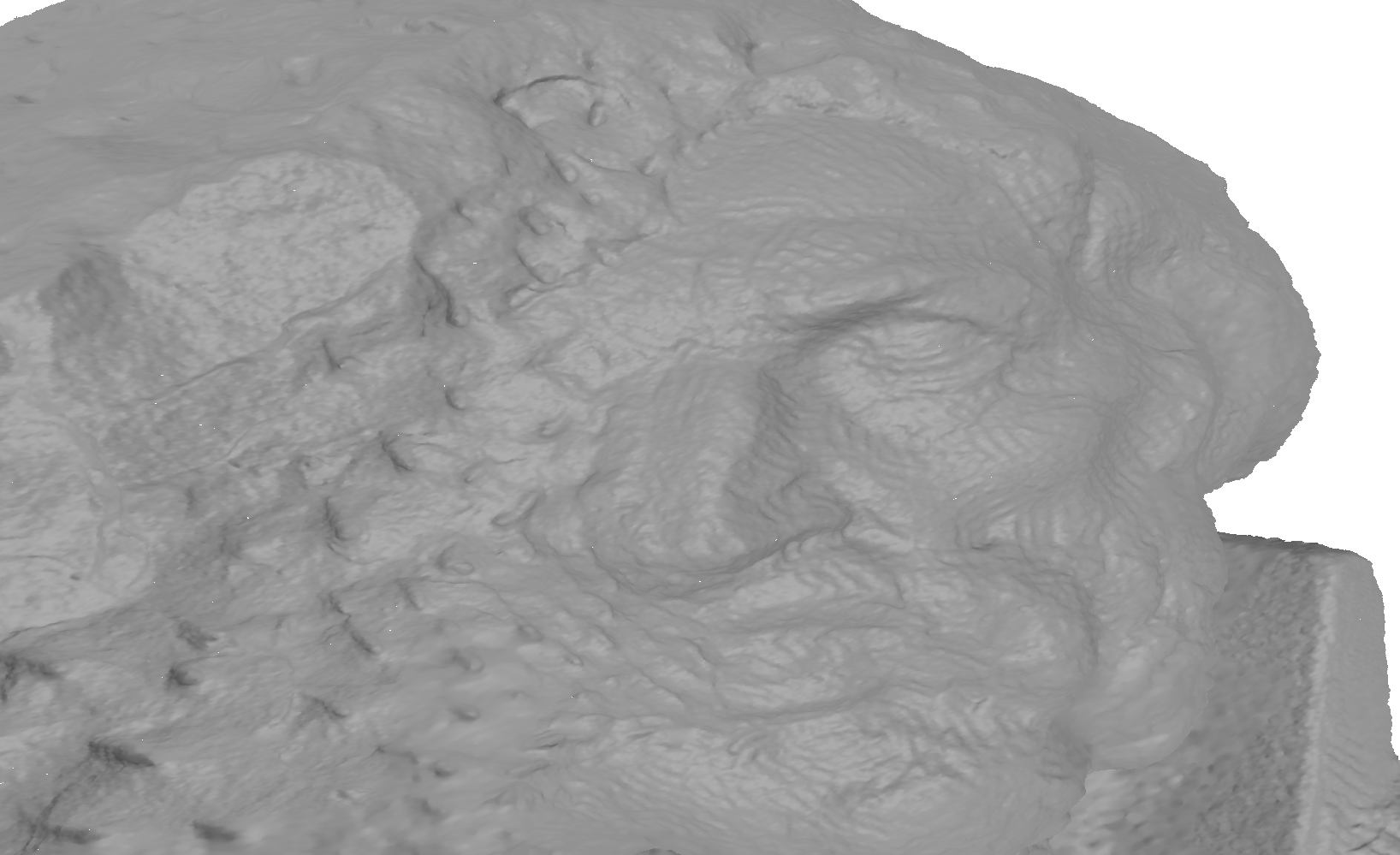}%
  \label{fig:stmathew_leaf_collapse_02_parent_ratio}
}
\hfil
\subfloat[St. Mathew, leaf collapse on, 0.25 point ratio.]{
  \includegraphics[width=.3\textwidth]{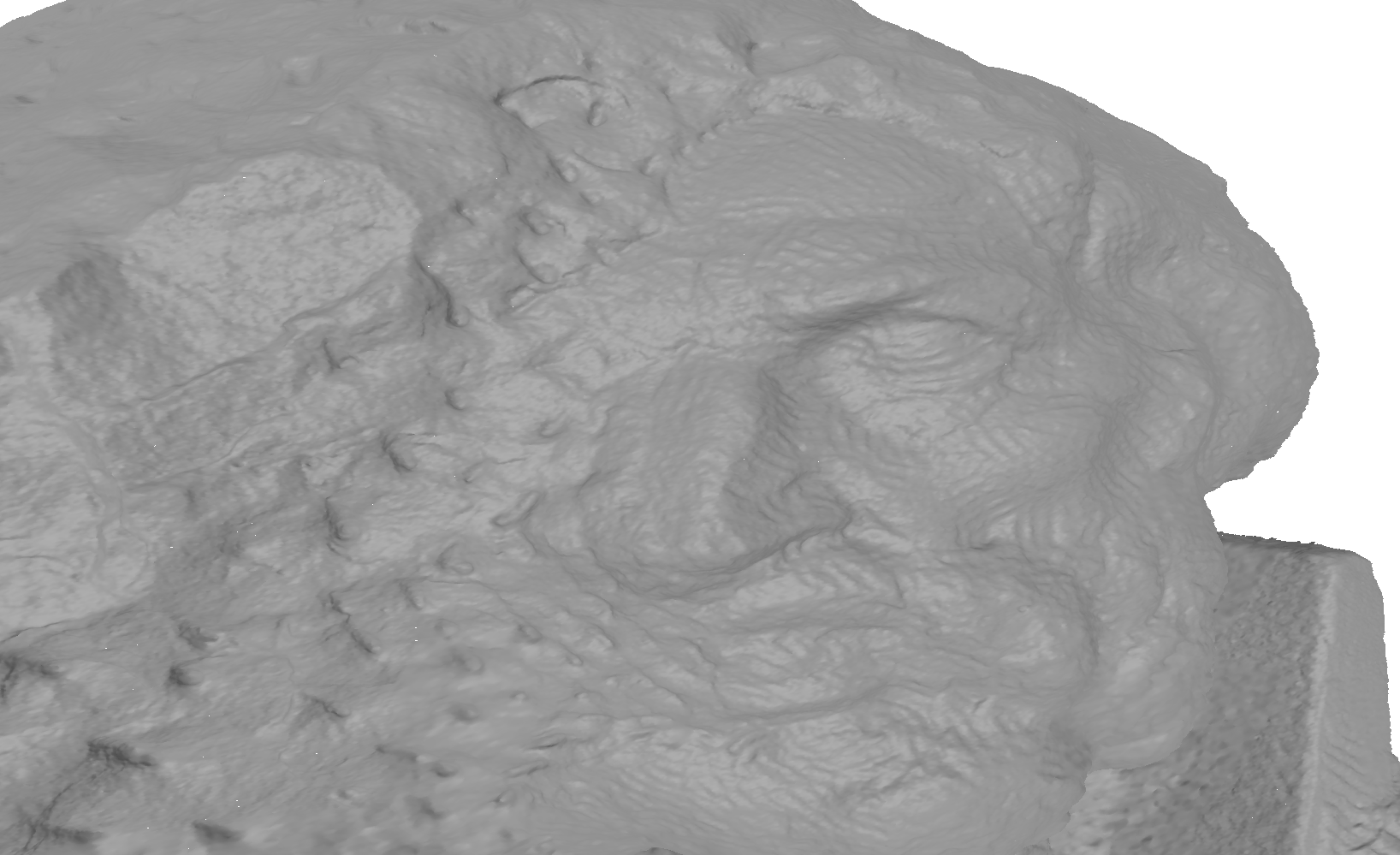}%
  \label{fig:stmathew_leaf_collapse_025_parent_ratio}
}
\hfil
\subfloat[St. Mathew, leaf collapse off, 0.25 point ratio.]{
  \includegraphics[width=.3\textwidth]{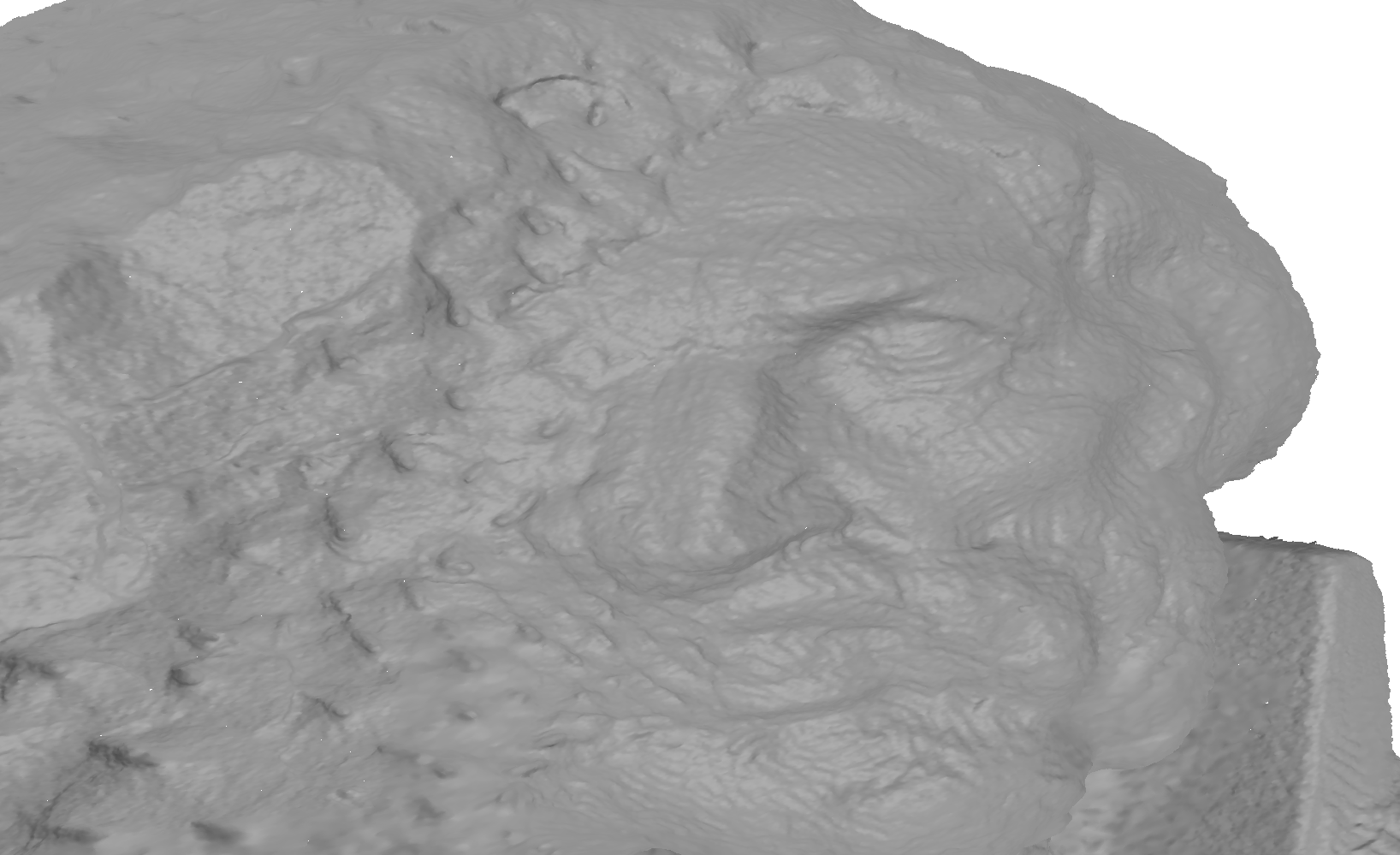}%
  \label{fig:stmathew_no_leaf_collapse_025_parent_ratio}
}
\hfil
\subfloat[Duomo, leaf collapse on, 0.2 point ratio.]{
  \includegraphics[width=.3\textwidth]{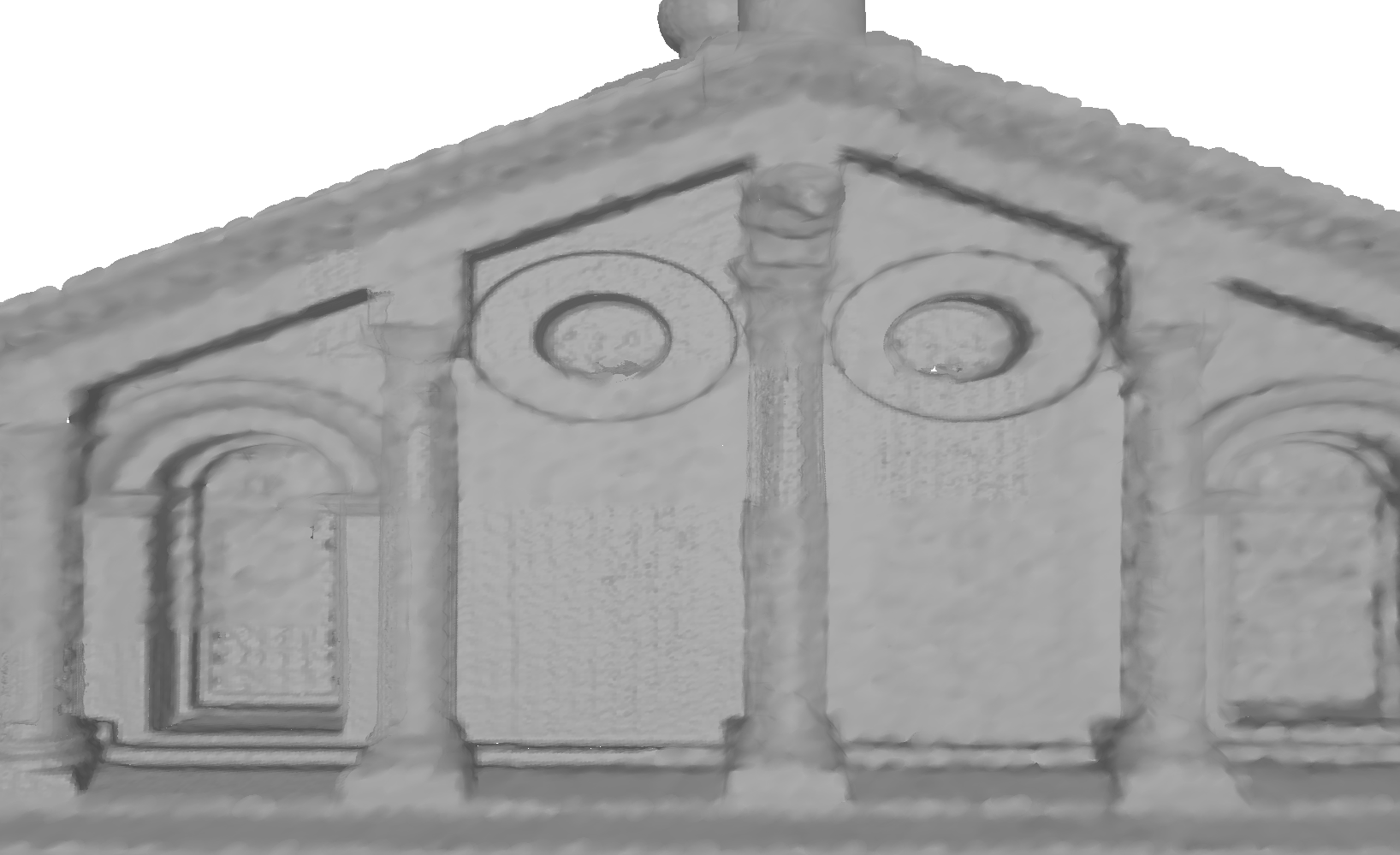}%
  \label{fig:duomo_leaf_collapse_02_parent_ratio}
}
\hfil
\subfloat[Duomo, leaf collapse on, 0.25 point ratio.]{
  \includegraphics[width=.3\textwidth]{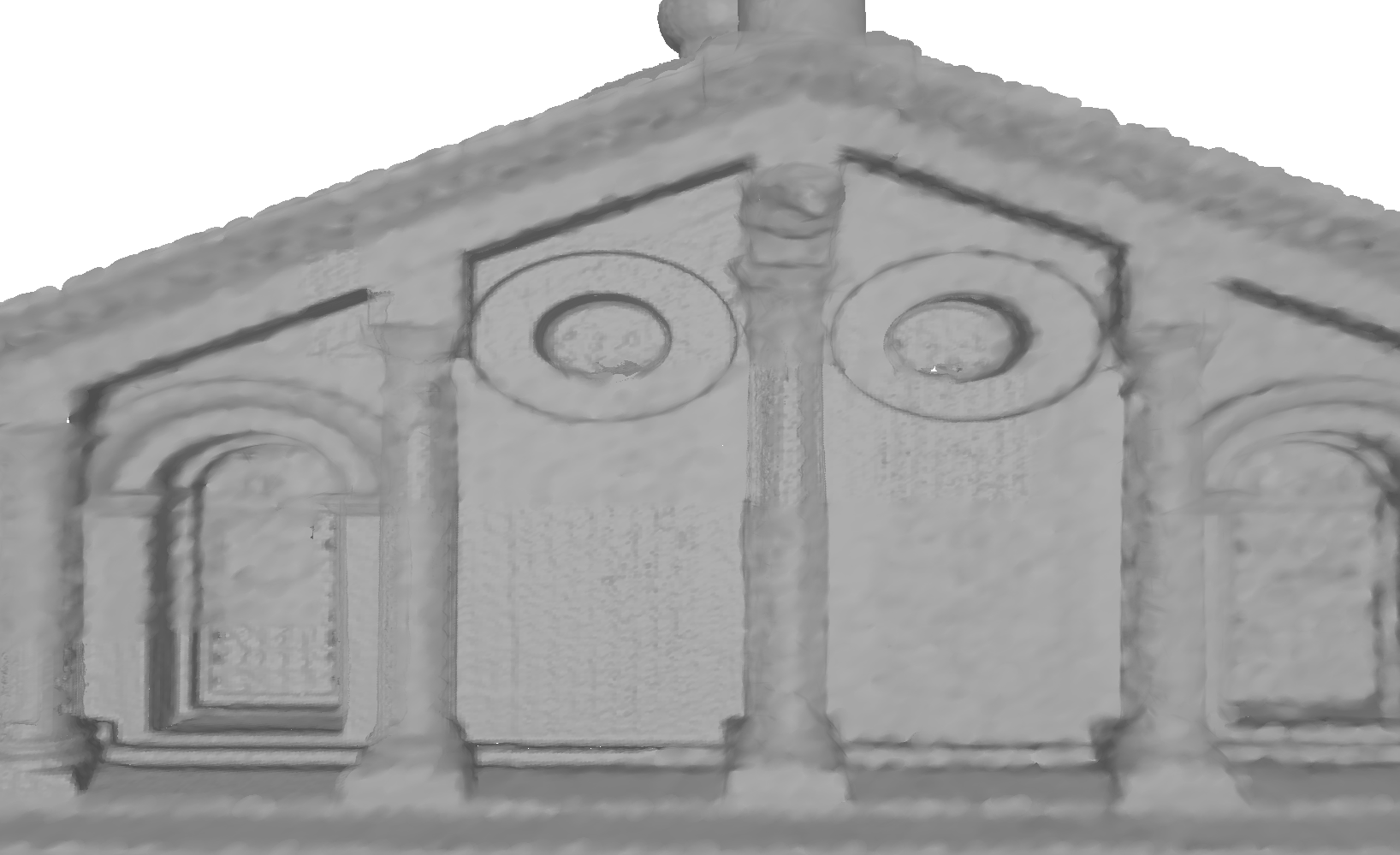}%
  \label{fig:duomo_leaf_collapse_025_parent_ratio}
}
\hfil
\subfloat[Duomo, leaf collapse off, 0.25 point ratio.]{
  \includegraphics[width=.3\textwidth]{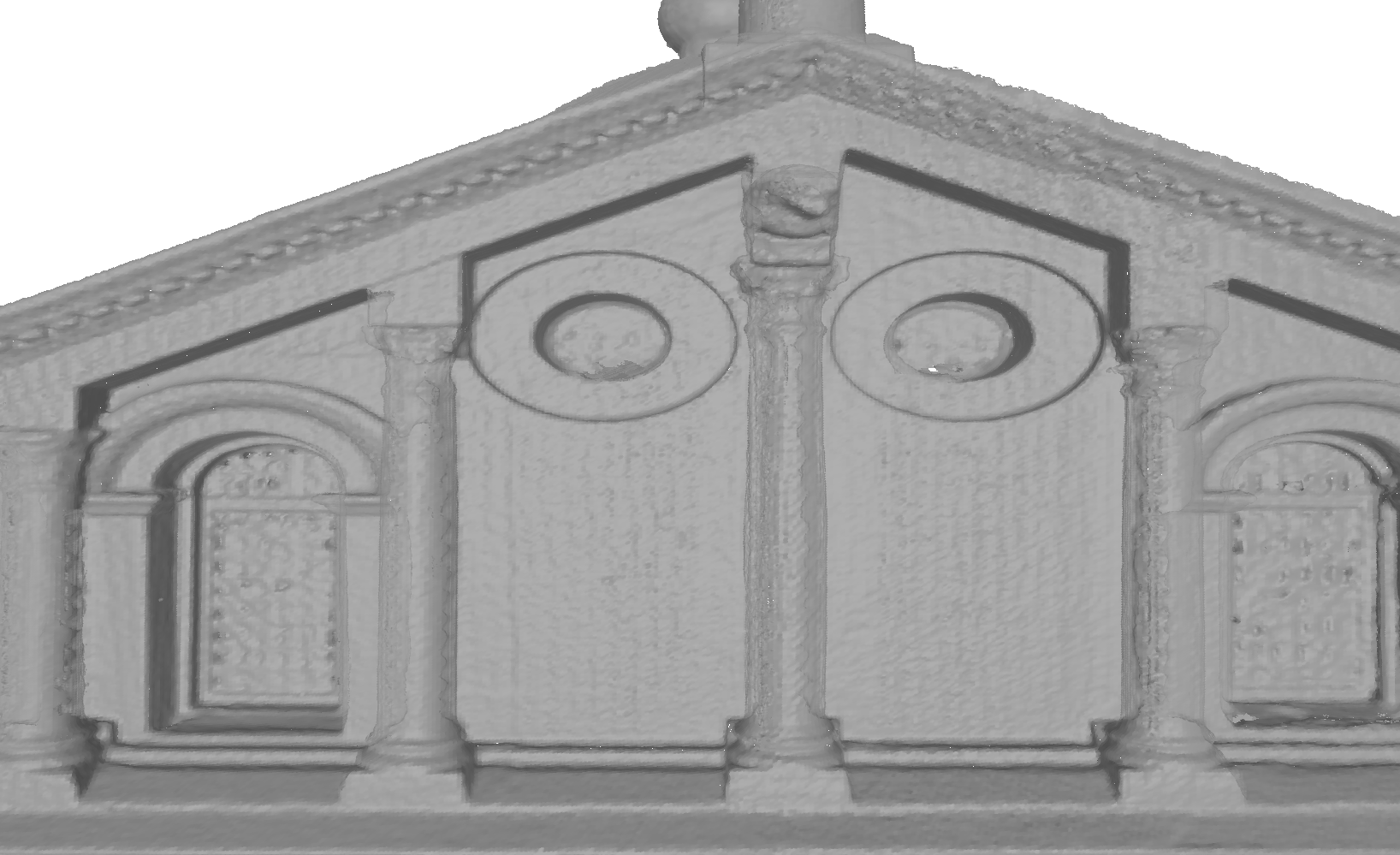}%
  \label{fig:duomo_no_leaf_collapse_025_parent_ratio}
}
\caption{Rendering comparison of hierarchies with different leaf collapse and parent to children point ratio parameters. As can be seen
from items (a) to (i), the final reconstructions are very detailed even at close range and the differences when the leaf
collapse is turned on are almost imperceptible for the David, Atlas and St. Mathew datasets. The hierarchy for Duomo suffers from lack
of density when leaf collapse is turned on because the dataset itself has smaller density in comparison with the others.}
\label{fig:leaf_collapse}
\end{figure*}

A second set of experiments were conducted to assess OMiCroN's behavior in terms of memory usage and performance. All experiments in this set read a sorted dataset directly from disk.  The test system had an Intel Core i7-6700 CPU, 16GB memory, NVidia GeForce GTX 1070 graphics card, and secondary SSD storage with roughly 130 MB/s reading speed.
Two main parameters impact OMiCroN's memory footprint: \textit{Leaf Collapse} optimization and parent to children point ratio, as shown in Table~\ref{table:leaf_collapse}. These also impact the reconstruction quality of the algorithm as can be seen in Figure \ref{fig:leaf_collapse}.




\begin{table}[]
\renewcommand{\arraystretch}{1.3}
\caption{Relationship between the algorithm reconstruction parameters -- leaf collapse, parent to children ratio -- and memory footprint, total hierarchy creation times, and
  average CPU usage per frame.}
\label{table:leaf_collapse}
\centering
\small
\begin{tabular}{|c|c|c|c|c|c|}	
\hline
Model & Coll & Ratio & Mem & Creation & CPU \\ [0.5ex]
\hline
David & On & 0.2 & 8.5GB & 146.3s & 7.6ms \\
David & On & 0.25 & 9.9GB & 151.2s & 8.8ms \\
David & Off & 0.2 & 21GB & 229.8s & 16.7ms \\
Atlas & On & 0.2 & 2.3GB & 77.8s & 11.9ms \\
Atlas & On & 0.25 & 3.0GB & 81.9s & 11.0ms \\
Atlas & Off & 0.2 & 11.5GB & 120.8s & 16.2ms \\
Mathew & On & 0.2 & 1.7GB & 59.6s & 13.7ms \\
Mathew & On & 0.25 & 2.2GB & 60.9s & 11.6ms \\
Mathew & Off & 0.2 & 8.4GB & 80.6s & 25ms \\
Duomo & On & 0.2 & 0.9GB & 31.0s & 18.2ms \\
Duomo & On & 0.25 & 1.2GB & 32.6s & 23.1ms \\
Duomo & Off & 0.2 & 4.5GB & 40.0s & 21.9ms \\
\hline
\end{tabular}
\end{table}

Even though limited to datasets that fit in RAM unless swap space is used, OMiCroN can be set up to fit a broad range of memory budgets.
For example, David originally occupies 11.2 GB in disk, while its maximum size in memory when using \textit{Leaf Collapse} is 8.5 or 9.9 GB, for parent to children point ratios of 0.2 and 0.25 respectively. In this case, a hierarchy with 0.2 ratio has memory usage of roughly 76\% of the original dataset size in disk. Values smaller than these are possible since reconstruction results shown in Figure~\ref{fig:leaf_collapse} are still acceptable. 
It is also important to note that the algorithm does not compress in any way the point or Morton code data. The use of such techniques would provide even better memory consumption. 

Table~\ref{table:leaf_collapse} also shows that the total hierarchy creation times and the average CPU usage per frame are affected by \textit{Leaf Collapse} optimization. The CPU times were obtained during a rendering session where the camera is constantly moving trying to focus the parts of the model being read from disk. 
For the David dataset, for example, it takes 88.2s to read the data from disk, while OMiCroN imposes an overhead ranging from 0.66 to 1.6 in the tested scenarios. We also notice that CPU times are probably affected by \textit{Leaf Collapse} optimization because the hierarchy is simplified when leaf nodes are removed, resulting in smaller hierarchy fronts.

The worklist size is the parameter that controls the work granularity in the hierarchy creation. In other words, it controls the throughput of new nodes available for the hierarchy creation threads to process. Table~\ref{table:work_list_size} shows the relationship between the worklist size and attributes that are expected to be directly affected by it.
It also shows that the front insertion delay scales linearly with the worklist size. As a consequence, larger worklists impose a longer delay for the user to see new parts of the cloud while navigating. Additionally, the optimal worklist size regarding front size is between 32 and 64. Since nodes are processed in a bottom-up manner and smaller fronts are expected to have nodes from shallower parts of the hierarchy, setups with smaller fronts are also expected to have processed more nodes from deeper levels than other setups with larger fronts, given the same time spent in processing. As a consequence, hierarchy construction time is reduced in setups with smaller fronts, as Table~\ref{table:work_list_size} also indicates. Similarly, benefits in overall performance of front evaluation are obviously related to smaller front sizes, resulting in less CPU overhead.

\begin{table}[]
\renewcommand{\arraystretch}{1.3}
\caption{Relationship between the worklist size and performance indicators:  front insertion delay, 
 front size,  hierarchy construction time and average CPU usage per frame. Numbers refer to the David dataset,
no leaf collapse and point ratio 0.25.}
\label{table:work_list_size}
\centering
\footnotesize
\begin{tabular}{|c|c|c|c|c|}
\hline
Worklist & Insertion & Front & Hierarchy & CPU \\ [0.5ex]
\hline
8 & 127ms & 529 & 274.8s & 19.5ms \\
16 & 212ms & 439 & 259.8s & 17.8ms \\
32 & 399ms & 401 & 248.6s & 16.0ms \\
64 & 831ms & 500 & 258.0s & 20.8ms \\
128 & 1646ms & 506 & 255.7s & 19.7ms \\
\hline
\end{tabular}
\end{table}


\subsection{Use cases}

We are also interested in evaluating OMiCroN's flexibility. To that end, we compared pipelines 2 and 3 from Figure~\ref{fig:pipelines}, i.e., constructing the hierarchy on-the-fly from a sorted point stream and reading a previously computed complete hierarchy file. The experiments were performed on the same machine as that used for the rendering latency tests (Section~\ref{section:rendering_tests}), but with more recent versions of  the dependency libraries and operating system. Pipeline 3 corresponds to the traditional approach, in which the hierarchy is read top-down in breadth-first order. This use case supports incremental visualization of the entire model from the beginning, starting with a coarse overview and progressively showing more details as the hierarchy is loaded.

\begin{figure}[!ht]
    \centering
    \includegraphics[width=\columnwidth]{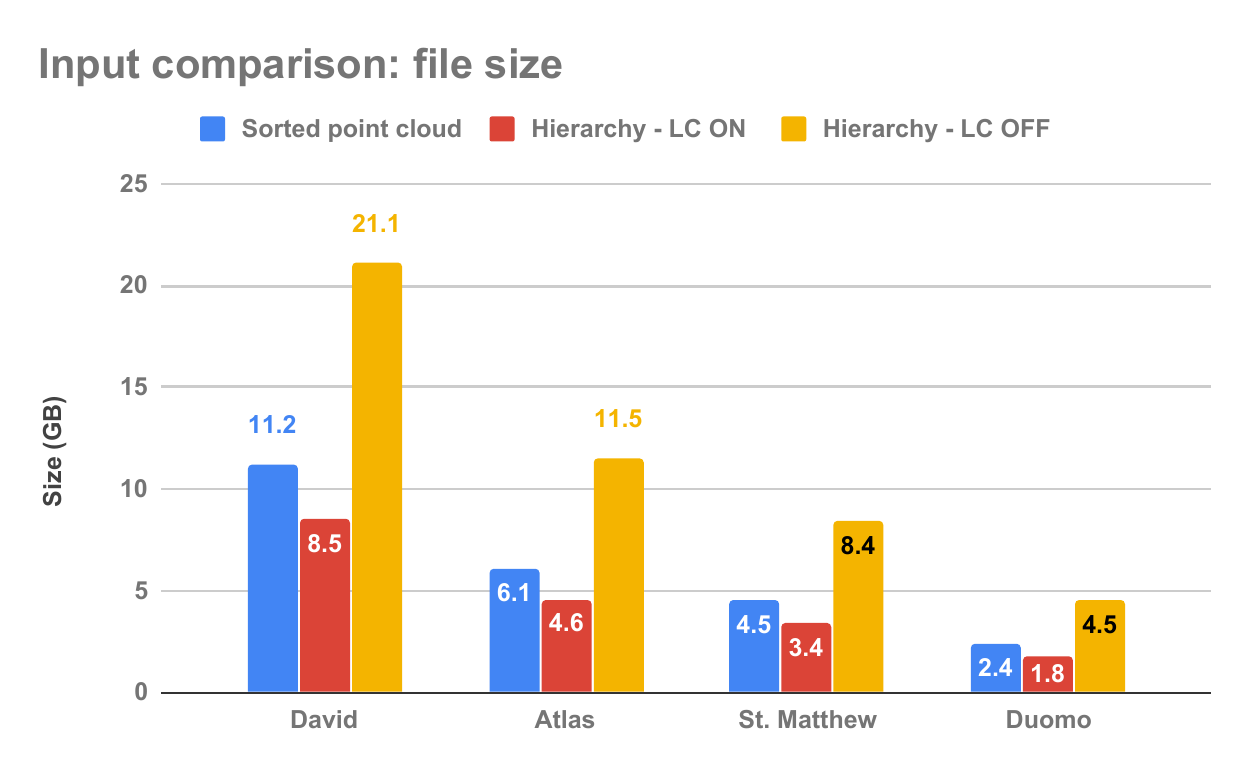}%
    \caption{Use case input file comparison. LC stands for Leaf Collapse. Saving a complete hierarchy file demands bigger storage than reading a sorted input cloud file and creating the hierarchy on-the-fly. The performance implications are depicted in Figure~\ref{fig:input_comparison_timings}.}
    \label{fig:input_comparison_file_size}
\end{figure}

\begin{figure}[!ht]
    \centering
    \includegraphics[width=\columnwidth]{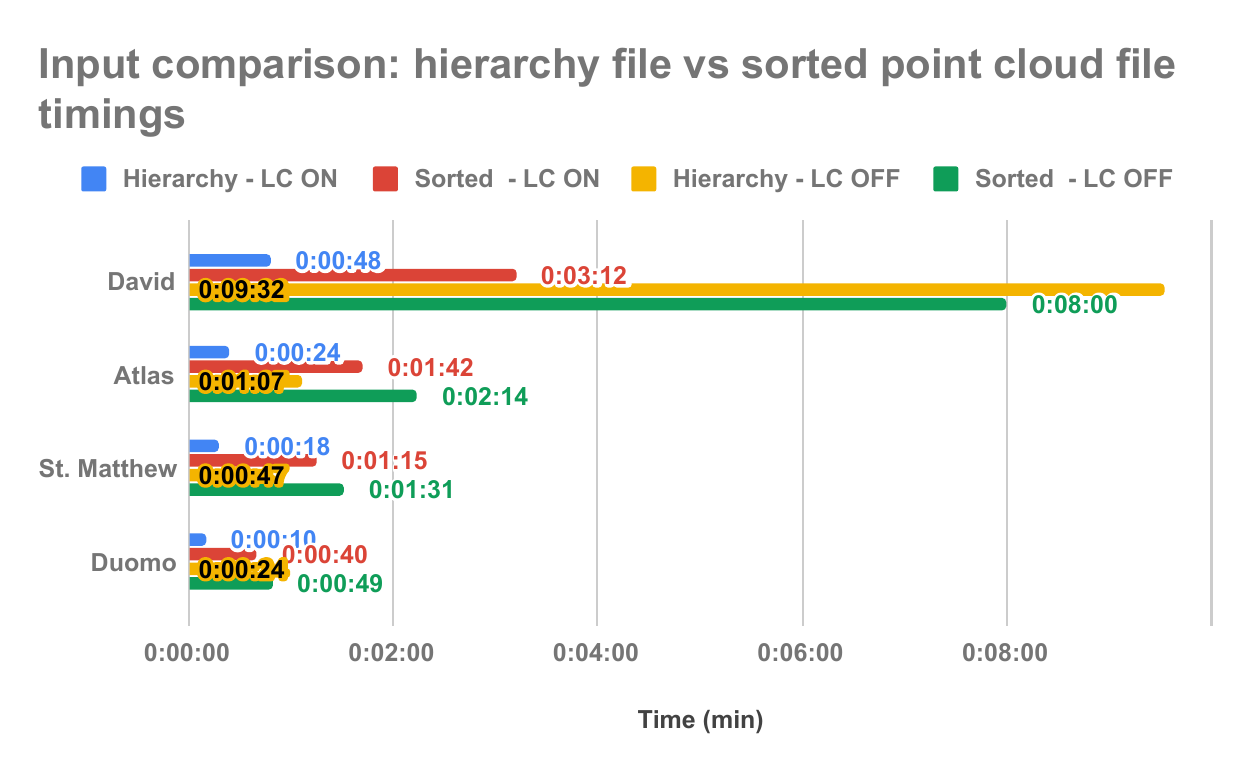}%
    \caption{Use case performance comparison. LC stands for Leaf Collapse. The values indicate the time needed to have the complete hierarchy in memory. Depending on the hierarchy file size it is better to build the hierarchy on-the-fly instead of loading it from disk because the same disk is used for reading and swap.}
    \label{fig:input_comparison_timings}
\end{figure}

Figure~\ref{fig:input_comparison_file_size} shows the comparison of input file size whereas Figure~\ref{fig:input_comparison_timings} compares the time needed to load a hierarchy file with the time needed to build the hierarchy from sorted point streams, as reported in Table~\ref{table:leaf_collapse}. Parallel rendering is enabled in all cases. 

The use case with the best performance depends on the hierarchy file size. For example, reading a hierarchy file with leaf collapse off for David has a significant performance penalty because the same disk is used to read a large file and for swap. The sorted list pipeline for this same case amortizes the swap overhead.


Even though reading the sorted list is generally slower than reading the hierarchy, there are important benefits to be taken into consideration, and the choice between one or the other would depend on the application scenario. First, the same sorted list can be used for creating the hierarchy with or without leaf collapse, and the leaf collapse parameters (compression level) can be chosen on demand. The second advantage is the reduced size of the input data, which leads to better performance for large clouds such as the David dataset without leaf collapse, as the swap overhead is reduced. The sorted list occupies roughly 50\% of the hierarchy file. While for a fast SSD disk this is less significant,  for other scenarios, such as network streaming, it could be very beneficial. Moreover, in the case of the hierarchy file, every different configuration generates a new file on disk. In other words, storing different hierarchies (e.g., with and without leaf collapse) is more wasteful than storing a single sorted list file. Of course, we could also store different sorted lists after performing leaf collapse, thus boosting the performance and reducing the space in disk, but for these tests we opted for storing the whole sorted list as we believe the extra flexibility is an important contribution.

\subsection{Comparisons}

We also found it useful to compare OMiCroN with other algorithms that create hierarchies for large datasets. To this end, we evaluated the hierarchy creation algorithm used in the large point cloud renderer Potree~\cite{Schutz2016}. The methodology was to compare the best cases in Figures~\ref{fig:partial_sort_mathew}, \ref{fig:partial_sort_atlas} and \ref{fig:partial_sort_david}, which include input, sorting, hierarchy creation and rendering, and the timings reported by Potree, which include input and hierarchy creation. All tests created hierarchies with depth 7.

\begin{figure}[!t]
    \centering
    \includegraphics[width=0.9\columnwidth]{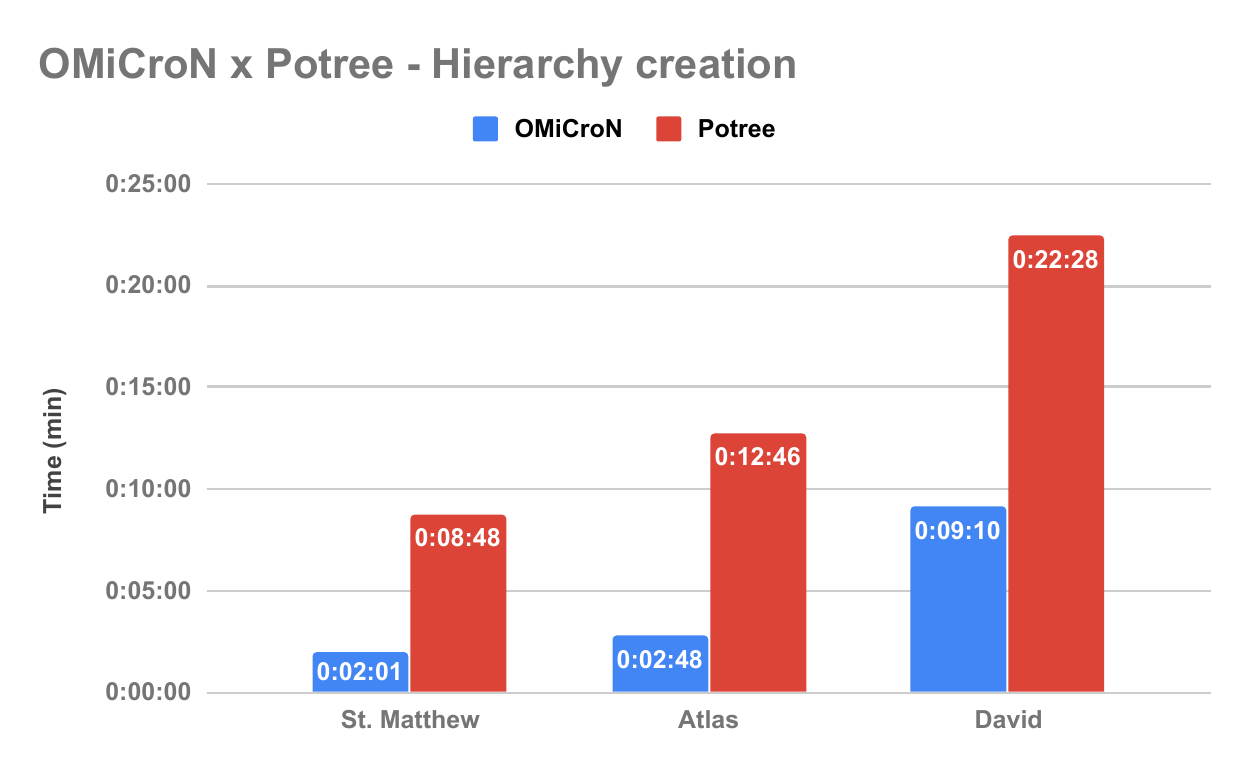}%
    \caption{OMiCroN and Potree\cite{Schutz2016} comparison. OMiCroN is more than 2 times faster for David and more than 4 times faster for St. Matthew and Atlas.}
    \label{fig:omicronXpotree}
\end{figure}

Figure~\ref{fig:omicronXpotree} shows the results for St. Matthew, Atlas and David. OMiCroN is more than 2 times faster for David and more than 4 times faster for St. Matthew and Atlas. An important detail is that Potree reports creating a hierarchy with only 68\% of the input points for David, whereas St. Matthew and Atlas result in 100\% of input points usage. OMiCroN is not only significantly faster, but its parallel renderer provides support for dataset inspection during the process, something that Potree cannot do. Please note that the desktop version of Potree was used for all comparisons.

 We also performed a comparison with the voxelization algorithm for large meshes described in~\cite{BAERT2014}. It should be noted that the algorithm operates on triangle meshes, and thus the input datasets are roughly twice as big as those containing only the vertices. However, since a voxelization is an abrupt simplification of the original dataset, the difference in input is compensated by the fact that Octree nodes handled by OMiCroN are populated with thousands or millions of points while the Octree nodes in the voxelization are boolean values, resulting in extremely compact Octrees with just a few KBytes. For example, the Octree generated by OMiCroN for the David without leaf collapse has more than 22GB. In our tests, \cite{BAERT2014} was given a memory quota of 16GB and set to a grid size of 128, which is equivalent to a hierarchy of depth 7. In our tests, OMiCroN finishes building the hierarchy 3 to 5 times faster than \cite{BAERT2014}, which indicates that, even in a traditional setup where preprocessing precedes rendering, OMiCroN is still very competitive. 

Finally, we performed a qualitative comparison with the incremental BVH construction algorithm proposed in \cite{Bittner2015}. The paper presents four versions of the algorithm: with or without global updates and with parallel searches or block parallel construction. Based on the presented results, we concluded that the BVH quality depends on three aspects: the version of the algorithm used, the stream ordering and the structure of the data. The same setup that results in a good BVH quality for a given dataset can result in a bad quality BVH for another one. Another conclusion is that it tends to perform better than non-incremental builders for simpler datasets with a lot of plane surfaces (Sibenik, Conference, Soda Hall, Pompeii, SanMiguel and PowerPlan) and worse for more complex, biological ones (Armadillo, Hairball and Happy Buddha). In the best case scenario (best algorithm choice, stream ordering and structure) the bvh quality can be up to 24\% better than top-down builders. In the worst case scenario (worst algorithm choice, stream ordering and structure) the quality can be up to 70\% worse.

To sum up, one should run the algorithm with several different setups to ensure a good quality BVH. Even in this case, the BVH can be worse than one created using a non-incremental builder because of the structure of the data. An incremental BVH construction algorithm probably could be changed to support parallel rendering, but the potential of generating a bad quality hierarchy could turn this option prohibitive. On the other hand, a non-incremental algorithm such as OMiCroN would construct a high quality octree regardless, providing rendering feedback in the process.



\section{Final remarks}
\label{sec:discussions}

In this work, we presented OMiCroN, a flexible and generic algorithm for rendering large point clouds. We know of no other method that can render incomplete hierarchies with full detail in parallel with its construction and data sorting. Rather, the vast majority of algorithms in this category rely on heavy preprocessing, which largely outweighs the time complexity of the rendering algorithm proper. OMiCroN, on the other hand, needs only a sorted prefix of the input geometry in Morton code order to start rendering. In practice, this sort can adapt to start rendering models as early as the time needed to read input. OMiCroN's feedback-based design allows construction of Octrees on-the-fly and can help implementors with accurate rendering feedback of the construction process. We also defined the novel idea of Hierarchy Oblique Cut, a strong concept that can be used to apply sweeps on hierarchies.

Additionally, OMiCroN opens the path for new workflows based on streaming of spatially sorted data. Supposing that large scans could be streamed directly in Morton order, the data could be rendered without any delays at all, enabling earlier detection of acquisition problems. Another advantage is that the hierarchical nature of Morton order can be explored, so datasets are sorted only once using a deep Morton code level but can be rendered by OMiCroN using a hierarchy with any level less or equal to the sorting level. This property renders the algorithm even more flexible, since a single sorted dataset can be used with many hierarchy setups.


\subsection{Out-of-core and incremental overview extension}

Even though in this work we have concentrated on describing our Oblique Cuts driven data structure, there are some possible extensions in order to generalize the method. An important improvement is to allow for an incremental version of OMiCroN. 

Briefly, in addition to oblique cuts, we could incorporate into OMiCroN also horizontal cuts that define depth intervals that could be used as loading units. 
Each horizontal cut would be constructed by the current version of OMiCroN, fed by an individual stream of points. Those streams could be constructed by sorting the input point cloud once in a deep hierarchy level and sampling the data with different granularities, since data ordered in a Morton curve of level $l$ is also ordered in any curve with level less than $l$.

Each horizontal cut could be restricted to a level interval by limiting the $fix$ operator bottom-up evaluation to a minimum predefined level. The entire structure would have a single rendering front and the horizontal cuts could be linked to leaves at front evaluation time, using an approach similar to placeholder substitution. This extension would also turn OMiCroN out-of-core by definition, since horizontal cuts could be released and constructed on-the-fly as needed.

\subsection{Future work}

Regarding future directions, OMiCroN has several possible paths to follow. Additionally to the implementation of the extended version, the splat renderer uses parameters set manually during the experiments, since it was not the focus of this work, rather we concentrated our efforts on the hierarchy construction and high-level rendering management. However, it could be further improved by developing methods to automatically find the optimal parameters, such as initial $u$ and $v$ vectors, and a better hierarchical representation of the splats~\cite{Wu2005}. Moreover, in theory, OMiCroN's deepest abstraction layer could be modified to use the algorithm in other Computer Graphics problems involving the use of Morton-ordered hierarchical structures, such as raytracing, voxelization and reconstruction.

\subsection{Source code}

OMiCroN's source code public repository can be found at \href{https://github.com/dsilvavinicius/OMiCroN}{https://github.com/dsilvavinicius/OMiCroN}.

\subsection{Acknowledgment}

This research was supported by the National Council for Scientific and Technological Development (CNPq). 

\bibliographystyle{cag-num-names}
\bibliography{omicron}

\end{document}